%% file: main.tex
\documentclass[letterpaper,journal,twoside]{IEEEtran}
\usepackage{amsmath,amsfonts,amssymb}
\usepackage{bm}
\usepackage{mathtools,esint}
\usepackage{algorithmic}
\usepackage{algorithm}
\usepackage{array}
\usepackage[caption=false,font=normalsize,labelfont=sf,textfont=sf]{subfig}
\usepackage{textcomp}
\usepackage{stfloats}
\usepackage{url}
\usepackage{siunitx}
\usepackage{verbatim}
\usepackage{graphicx}
\usepackage{cite}
\usepackage{pgf,pgfplots,tikz,ifthen}
\usetikzlibrary{3d,decorations.pathreplacing,shapes,arrows,arrows.meta,positioning,intersections,calc,spy,perspective,patterns,patterns.meta}
\usepackage{xparse}
\pgfplotsset{compat=1.18}
\usepackage{adjustbox}
\usepackage[nolist,nohyperlinks]{acronym}
\usepackage{orcidlink}
\usepackage{threeparttable}

\newcommand{\Lpar}[1]{\!\left(#1\right)}

\newcommand{\Lnorm}[1]{\!\left\lVert#1\right\rVert}

\newcommand\copyrighttext{%
   \footnotesize
   \textcopyright 2024 IEEE.
   Personal use of this material is permitted. Permission from IEEE must be obtained for all other uses, in any current or future media, including reprinting/republishing this material for advertising or promotional purposes, creating new collective works, for resale or redistribution to servers or lists, or reuse of any copyrighted component of this work in other works.
}
\newcommand\copyrightnotice{%
\begin{tikzpicture}[remember picture,overlay]
\node[anchor=south,yshift=5pt] at (current page.south) {\fbox{\parbox{\dimexpr\textwidth-\fboxsep-\fboxrule\relax}{\copyrighttext}}};
\end{tikzpicture}%
}

\begin{document}

\begin{acronym}[MUSHRA]
\acro{VR}{virtual reality}
\acro{AR}{augmented reality}
\acro{ASR}{automatic speech recognition}
\acro{GA}{geometrical acoustics}
\acro{ISM}{Image Source Method}
\acro{RTM}{Ray Tracing Method}
\acro{ART}{Acoustic Radiance Transfer}
\acro{RARE}{Room Acoustics Rendering Equation}
\acro{RIR}{room impulse response}
\acro{FIR}{finite impulse response}
\acro{FDN}{Feedback Delay Network}
\acro{ARN}{Acoustic Rendering Network}
\acro{DWN}{Digital Waveguide Network}
\acro{DWM}{Digital Waveguide Mesh}
\acro{WGW}{Waveguide Web}
\acro{SDN}{Scattering Delay Network}
\acro{NED}{normalized echo density}
\acroplural{NED}[NEDs]{normalized echo densities}
\acro{EDC}{energy decay curve}
\acro{EDT}{early decay time}
\acro{BRDF}{bidirectional reflectance distribution function}
\acro{HRTF}{head-related transfer function}
\acro{MUSHRA}{Multiple Stimulus with Hidden Reference and Anchor}
\acro{BRAS}{Benchmark for Room Acoustical Simulation}
\acro{VBAP}{vector-based amplitude panning}
\acro{HOA}{higher-order ambisonics}
\end{acronym}


\title{Room Acoustic Rendering Networks with\\Control of Scattering and Early Reflections}

\author{Matteo Scerbo\,\orcidlink{0000-0001-6057-7432},
\and
Lauri Savioja,~\IEEEmembership{Senior member,~IEEE}\,\orcidlink{0000-0002-8261-4596},
\and
Enzo De Sena,~\IEEEmembership{Senior member,~IEEE}\,\orcidlink{0000-0002-8007-4370}
\thanks{This work was supported by the Engineering and
Physical Sciences Research Council under the SCalable Room Acoustics Modelling Grant EP/V002554/1.}
\thanks{Matteo Scerbo and Enzo De Sena are with the Institute of Sound Recording, University of Surrey, GU2 7XH Guildford, U.K. (e-mail: m.scerbo@surrey.ac.uk; e.desena@surrey.ac.uk).}
\thanks{Lauri Savioja is with the Department of Computer Science, Aalto University, 02150 Espoo, Finland (e-mail: lauri.savioja@aalto.fi).}
}

\markboth{SUBMITTED TO IEEE/ACM Transactions on Audio, Speech, and Language Processing}%
{Scerbo \MakeLowercase{\textit{et al.}}: Room Acoustic Rendering Networks with Control of Scattering and Early Reflections}


\maketitle
\copyrightnotice

\input{sections/abstract}

\input{sections/introduction}

\input{sections/background}

\input{sections/proposed}

\input{sections/results}

\input{sections/conclusions}

\section*{Acknowledgments}
We thank Sebastian J. Schlecht for the useful discussions on lossless matrices, and Randall Ali for the useful discussions on notation and presentation of the geometrical aspects.

\input{sections/appendix}

\bibliographystyle{IEEEbib}
\bibliography{references}

\input{biography}

\vfill

\end{document}

%% file: sections/abstract.tex
\begin{abstract}
Room acoustic synthesis can be used in \acf{VR}, \acf{AR} and gaming applications to enhance listeners' sense of immersion, realism and externalisation.
A common approach is to use \acf{GA} models to compute impulse responses at interactive speed, and fast convolution methods to apply said responses in real time.
Alternatively, delay-network-based models are capable of modeling certain aspects of room acoustics, but with a significantly lower computational cost.
In order to bridge the gap between these classes of models, recent work introduced delay network designs that approximate \ac{ART}, a \ac{GA} model that simulates the transfer of acoustic energy between discrete surface patches in an environment.
This paper presents two key extensions of such designs.
The first extension involves a new physically-based and stability-preserving design of the feedback matrices, enabling more accurate control of scattering and, more in general, of late reverberation properties.
The second extension allows an arbitrary number of early reflections to be modeled with high accuracy, meaning the network can be scaled at will between computational cost and early reverberation precision.
The proposed extensions are compared to the baseline \ac{ART}-approximating delay network as well as two reference \ac{GA} models.
The evaluation is based on objective measures of perceptually-relevant features, including frequency-dependent reverberation times, echo density build-up, and early decay time.
Results show how the proposed extensions result in a significant improvement over the baseline model, especially for the case of non-convex geometries or the case of unevenly distributed wall absorption, both scenarios of broad practical interest.
\end{abstract}

\begin{IEEEkeywords}
Room acoustics modeling, \aclp{FDN}, \acl{RARE}.
\end{IEEEkeywords}

%% file: sections/introduction.tex
\section{Introduction}
\label{sec:introduction}
\IEEEPARstart{T}{he} field of room acoustics modeling is concerned with predicting the acoustical behaviour of environments.
This finds application in a wide variety of areas, ranging from architecture and civil engineering -- where designers need to account for speech intelligibility and noise levels in a space before it is constructed -- to movies, video-games, and \ac{VR} / \ac{AR} applications -- where reverberation can enhance listeners’ sense of realism~\cite{apostolopoulos2012road}, immersion~\cite{potter2022relative},  and externalization~\cite{geronazzo2020minimal}.
In many situations, it is desirable for the simulation to be performed in real-time, updating interactively while sound sources and receivers move in the virtual environment.

Wave-based room acoustics models are based on the physical laws governing sound wave propagation, and can therefore account for all acoustic phenomena.
This comes at the cost of considerable computational complexity, making such models unviable to run in real-time for broad-band signals.
\Acf{GA} models approximate the propagation of sound as being characterized by rays, as opposed to waves.
This approximation cannot correctly capture phenomena such as room resonances and modal behaviour~\cite{fifty_years}, a problem that can be addressed by considering hybrid wave-based/\ac{GA} models~\cite{southern2013room}.
Similar to other classes of models, \ac{GA} ones do not result in ``authentic'' reproduction, in the sense that the difference between simulations and measurements of the same scene are clearly audible~\cite{brinkmann2019round}.
Crucially, however, they are capable of rendering acoustics in a manner that is sufficiently plausible, a more important criterion when rendering for video-games, music production and VR/AR~\cite{brinkmann2019round,fifty_years, fifty_more}.

Even with the fastest \ac{GA} models, only a limited number of reflections can be simulated in real time with limited computational resources.
The high reflection orders which characterize an environment's perceived reverberation time lead to steadily increasing computational load.
The predicted \acfp{RIR} are most often applied to the source signal using convolution with both of the mentioned model classes.
This yet increases the computational requirements -- the computational cost of convolution scales with the duration of the \ac{RIR} -- and can introduce undesirable latency~\cite[Chap.~15]{vorlander}.

A different way of applying reverberation to a signal is through the use of recursively connected delay lines.
This characterizes the delay-based class of models, originating with the Schroeder reverberator~\cite{schroeder_colorless, schroeder_natural}, with \acfp{FDN}~\cite{jot_fdn, rocchesso_fdn} being the most widespread.
These methods are more computationally efficient than any other, but do not intrinsically constitute physical models of an environment.
Recent work~\cite{arn, fdn_art, efficient_sdn, das2023grouped, atalay2022scattering} proposed room acoustics models that exploit the advantages of each method.
These consist of delay-based structures
while incorporating \ac{GA} components in their design, thus providing fast modeling based on physical attributes of the space.
\acfp{ARN}~\cite{arn} segment the room surface into patches, modeling the propagation of energy between them.
\acs{FDN}-\acs{ART}~\cite{fdn_art} employs a small number of delay lines, based on the most prominent dimensions of the space.
\acfp{SDN}~\cite{efficient_sdn, atalay2022scattering} are designed around first-order reflection paths.
Due to their relatively recent emergence, this class of models still suffers from some limitations with respect to convolution-based \ac{GA}.
They lack control of scattering coefficients, an important acoustical parameter, and can only produce reliably accurate reflections for the first order.

Other works made joint use of \ac{GA} and delay-based models, either by employing \ac{GA} for early reflections and a delay-based model for late reverberation~\cite{diva} or by partitioning the space to be modeled, computing responses for all partitions, and connecting the resulting ``convolution blocks'' with delay lines~\cite{virtualacoustics}.
The focus of this paper is on methods that incorporate physically relevant information directly in a delay-based structure, to maximize physical accuracy and reliability while maintaining the efficiency of such models.

In this paper, we propose two key extensions of previous \acs{GA}-oriented, delay-based models.
Following work by Bai~\textit{et~al}.~\cite{arn}, our extensions rely on the \ac{RARE} -- the analytical formulation underlying all \ac{GA} models.
First, we introduce two novel feedback matrix designs (``Sinkhorn-Knopp'' design, ``uniform'' design), offering more flexible control of scattering for late reverberation.
The second extension (high-order injection) enables the early response to be scaled between accuracy and computational efficiency, by selecting the number of reflection orders to be modeled with high fidelity outside of the recursive network.
These extensions facilitate the modeling of environments with non-convex geometries or unevenly distributed wall absorption, as is often needed for practical applications.

Section~\ref{sec:background} reviews the \ac{RARE} and its discretization, the fundamental concepts of delay-network structures, and prior art delay-network-based models approximating the \ac{RARE}.
Section~\ref{sec:proposed} details the construction and limitations of our proposed method, and Section~\ref{sec:results} illustrates its performance in several validation tests.
Finally, Section~\ref{sec:conclusions} presents concluding remarks and discusses future work.

%% file: sections/background.tex
\section{Background}
\label{sec:background}

This section will outline the fundamentals of radiance-based acoustic modeling and the \acf{ART} method.
Then, it will illustrate the similarities and connections between \ac{ART} and delay-network-based models.
Finally, it will describe matrix design strategies employed in the prior art.

\subsection{Geometrical Acoustics}

As previously mentioned, \ac{GA} models rely on acoustic rays to characterize sound propagation.
The underlying mathematical model has its roots in geometrical optics~\cite[Chap~2]{form_factors}.
When applied to acoustic energy propagating in rooms, said model takes the form of the \acf{RARE}~\cite{rare}, which unfortunately cannot be solved analytically in the general case.
In practice, \ac{GA} methods, such as ray tracing or beam tracing, act as numerical solvers for this equation, with each model in the class being characterized by different approximations and/or discretizations.

Table~\ref{tab:variables} lists the notation\footnote{This paper uses a notation similar but not identical to~\cite{rare} and is meant to emphasise the links with delay-network-based room acoustic models.} and terminology used in this paper for physical measures.
The remainder of this subsection will describe radiance, the concept upon which the \ac{RARE} is based on, followed by definitions of the \ac{RARE} and \ac{ART}.

\begin{table}[tb]
    \begin{minipage}{\columnwidth}
    \caption{\itshape Notation of the physical quantities used in this paper.}
    \centering
    \renewcommand{\arraystretch}{1.2}
    \begin{tabular}{|c|c|c|}
    \hline
    Notation & Units & Nomenclature \\
    \hline
    \hline
    $P$ & \si{\watt} & Power \\
    \hline
    $\Omega, d\omega$ & \si{\steradian} & Solid angle \\
    \hline
    $A, dA$ & \si{\meter\squared} & Area \\
    \hline
    $A_\text{p}, dA_\text{p}$ & \si{\meter\squared} & Projected area \\
    \hline
    $I$ & \si{\watt\per\meter\squared} & Sound intensity\footnote{In the context of the \ac{RARE}, sound intensity corresponds to the radiometric quantity called ``irradiance'' or ``radiosity''. This is not to be confused with ``radiant intensity'', a radiometric quantity measured in \si{\watt\per\steradian}.} \\
    \hline
    $L$ & \si{\watt\per\steradian\per\meter\squared} & Radiance \\
    \hline
    $\rho$ & \si{\per\steradian} & \acs{BRDF} \\
    \hline
    $x, x', x''$ & \si{\meter} & Position vectors \\
    \hline
    $v, n$ & N.A. & Direction vectors \\
    \hline
    \end{tabular}
    \label{tab:variables}
    \end{minipage}
\end{table}

\subsubsection{Radiance}

\begin{figure}[t]
    \centering
    \begin{adjustbox}{max width=0.5\textwidth}
    \input{figures/points_on_surfs}
    \end{adjustbox}
    \caption{Examples of directional vectors related to three boundary points.}
    \label{fig:points on surfs}
\end{figure}
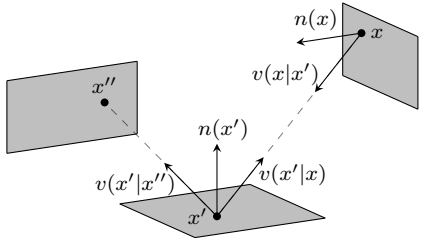

Radiance, denoted by $L$, can be intuitively interpreted as energy traveling along the path of a ray -- from a given point, in a given direction~\cite{rare}.
We will use the notation $v(x|x')$ for the directional vector based in $x$ and pointed towards $x'$ (see \figurename~\ref{fig:points on surfs}), therefore $L \Lpar{x, v(x|x'), t}$ will denote the radiance emitted from point $x$ in the direction of $x'$ at time~$t$.
Formally, radiance is defined as the derivative of power over solid angle and projected area.
For example, the power radiated from surface $A_1$ to surface $A_2$ at time $t$ is~\cite{rare}
\begin{equation}
    P \Lpar{A_1, A_2, t}
    =
    \iint\displaylimits_{A_2}
    \iint\displaylimits_{A_1}
    L \Lpar{x, v(x|x'), t}
    d\omega_{x}(x')
    dA_{\text{p}\, x'}(x)
    \, ,
    \label{eq:radiance definition integral}
\end{equation}
where
$dA_{\text{p}\, x'}(x)$ is the differential area $dA(x)$ projected towards $x'$, and $d\omega_{x}(x')$ is the differential solid angle of the opening $dA(x')$ observed from $x$.
Explicit expressions for these differential terms are reported in the Appendix. 

\subsubsection{\acl{RARE}}

The \ac{RARE} is a Fredholm integral equation of the second kind, characterizing the radiance between points on the entire reflective surface~$A$ of an environment.
It is defined in~\cite{rare}, from which all the following remarks are drawn.
In other words, given an environment defined by its boundary surface, the radiance reflected by each surface point is defined by the \ac{RARE} as an integral sum of radiance contributions from all other surface points.
This provides a recursive expression for the propagation of energy within the environment, over time.
The \ac{RARE} defines the radiance outgoing from the point $x'$ in the direction of $x''$ at time $t$ as
\begin{align}
    L \Lpar{x', v(x'|x''), t}
    = & \ \nonumber
    \iint\displaylimits_A
    L \Lpar{x, v(x|x'), t}
    \\ \nonumber
    & \hphantom{=\iint\displaylimits_A}
    *
    R \Lpar{x, x', v(x'|x'')}
    dA(x)
    \\ & \ +
    L_0 \Lpar{x', v(x'|x''), t}
    \, ,
    \label{eq:RARE}
\end{align}
where $*$ denotes convolution (the $R$ operator includes a time delay), and $L_0$ is the radiance originated\footnote{Notation in radiometry uses $L_0$ for the radiance being emanated by the surface -- not reflected or transmitted. This is not relevant in acoustics, where it is common to separate energy sources from the reflecting boundary, and $L_0$ is instead considered to be the primary reflected radiance. See the Appendix for more details.} by the surface.
The term $R\Lpar{x, x', v(x'|x'')}$ is the so-called reflection kernel:
\begin{align}
    \nonumber
    R \Lpar{x, x', v(x'|x'')}
    = & \ 
    \rho \Lpar{x'|\, v(x'|x), v(x'|x'')}
    \\ & \ 
    \times
    G(x, x')
    V(x, x')
    D(x, x')
    \, ,
    \label{eq:refl kernel}
\end{align}
where $\rho$ models the scattering and energy losses (due to material absorption) at $x'$, $G$ models the scaling of energy due to the distance and arrangement of $dA(x)$, $dA(x')$, $V$ models the visibility between $x$, $x'$, and $D$ models the propagation delay and air absorption between $x$, $x'$.
See the Appendix for the explicit definitions of these terms, and a discussion of how sources and receivers are modeled in the \ac{RARE}.

\subsubsection{\acl{ART}}

As previously mentioned, all \ac{GA} methods constitute numerical solvers for the \ac{RARE}.
In particular, the \acf{ART} method~\cite{td-art} evaluates the integral in Eq.~\eqref{eq:RARE} by segmenting the entire reflecting surface $A$ into $N$ discrete patches $A_1, A_2, \hdots, A_N$, and averaging the transferred radiance and reflection kernels over them.
For example, the discretized surface radiance traveling from patch $i$ to patch $j$ is defined by integrating the continuous radiance over $A_i$ and $A_j$.
Time $t$ is also discretized in samples, $n$.
The discretized surface radiance at time sample $n$ is defined by sampling the continuous radiance at time ${t_n = n \Delta t}$, where $\Delta t$ is the discrete time step.
Then, the discrete radiance $\hat{L}$ can be defined with respect to the continuous radiance $L$ as
\begin{align}
    \hat{L}_{i \to j} (n)
    = & \ 
    \iint\displaylimits_{A_j}
    \iint\displaylimits_{A_i}
    L \Lpar{x, v(x|x'), n \Delta t}
    \frac{dA(x)}{A_i}
    \frac{dA(x')}{A_j}
    \, .
    \label{eq:discretized radiance}
\end{align}
Note that ${\iint_{A_i} f(x) \frac{dA(x)}{A_i}}$ is the average value of $f$ over~$A_i$.
The reflection kernel is expressed by a matrix $\bm{\hat{R}}$, with elements defined as
\begin{align}
    \hat{R}_{h \to i, i \to j}
    = & \ 
    \iint\displaylimits_{A_j}
    \iint\displaylimits_{A_i}
    \iint\displaylimits_{A_h}
    R \Lpar{x, x', v(x'|x'')}
    \label{eq:discretized kernel}
    \\ \nonumber
    & \hphantom{=
    \iint\displaylimits_{A_j}
    \iint\displaylimits_{A_i}
    \iint\displaylimits_{A_h}} \times
    dA(x)
    \frac{dA(x')}{A_i}
    \frac{dA(x'')}{A_j}
    \, ,
\end{align}
which denotes the reflection kernel ${h \to i \to j}$, i.e.\@ from patch $h$ to patch $j$ through patch $i$.
The choice of subscript notation for $\hat{R}_{h \to i, i \to j}$ is meant to emphasise how the operator connects two paths, ${h \to i}$ and ${i \to j}$, which will be later modeled as delay lines.
The indexing convention of $\bm{\hat{R}}$ is such that each row is associated with one of these paths and so is each column -- the exact ordering is inconsequential.
Note that the discretization of $\bm{\hat{R}}$ also involves a temporal discretization of the delay operators, left implied in~\eqref{eq:discretized kernel}.

The integral in Eq.~\eqref{eq:RARE}, now discretized, is a sum:
\begin{align}
    \hat{L}_{i \to j} (n)
    = & \ 
    \sum_h^{N}
    \hat{L}_{h \to i} (n)
    \circledast
    \hat{R}_{h \to i, i \to j}
    +
    \hat{L}_{0, i \to j} (n)
    \, ,
    \label{eq:discretized RARE}
\end{align}
where $\circledast$ denotes discrete-time convolution, $\hat{L}_{0, i \to j} (n)$ is the discretized primary reflected radiance, and $\sum_h^{N}$ is a summation over all $N$ patches.

It should be noted that the convolution ${\hat{L}(n) \circledast \hat{R}}$ in Eq.~\eqref{eq:discretized RARE} results in temporal spreading, since the integration of the underlying delay operator $D(x, x')$ involves different delay values.
In the simplest case, this operation is modeled as a scaling coefficient and a single (integer) delay, which is a reasonable approximation when the surface patches are small compared to the distance between them.
Defining~$\bm{D}$ as the vector of time delays in~$\bm{\hat{R}}$, and~$\bm{\hat{S}}$ as the matrix of scaling coefficients of~$\bm{\hat{R}}$, Eq.~\eqref{eq:discretized RARE} takes the simplified form
\begin{align}
    \hat{L}_{i \to j} (n)
    = & \ 
    \sum_h^{N}
    \hat{L}_{h \to i} (n - D_{h \to i})
    \hat{S}_{h \to i, i \to j}
    +
    \hat{L}_{0, i \to j} (n)
    \, .
    \label{eq:ART as FDN}
\end{align}
Note that, due to the energy-conserving property~\cite{brdf} of physically realistic \acp{BRDF} $\rho$, the matrix $\bm{\hat{S}}$ always satisfies the condition
\begin{equation}
    \sum_h \hat{S}_{h \to i, i \to j}
    \le 1 \ \forall i, \forall j
    \, .
    \label{eq:R energy conservation}
\end{equation}
The methods proposed in Section~\ref{sec:proposed-matrices} hinge on this property.

Finally, we define the \emph{direct-incidence}, \emph{injection}, and \emph{detection} operators.
The latter two are sometimes referred to as \emph{initial tracing} and \emph{final gathering}, respectively, and are typically computed using ray tracing.
The injection operators, $\hat{s}_{i \to j}$, define the relationship between the source intensity and the primary reflected radiance:
\begin{align}
    \hat{L}_{0, i \to j} (n)
    = & \ 
    \hat{s}_{i \to j}\, I_s (n)
    \, ,
    \label{eq:ART injection}
\end{align}
where $I_s (n)$ is the input sound intensity at a source point $x_s$.

The detection operators, $\hat{r}_{h \to i}$, define the relationship between the discretized radiance values $\hat{L}_{h \to i} (n)$ and the corresponding intensity contributions at a receiver point $x_r$, which make up the output sound intensity $I_r (n)$:
\begin{equation}
    I_r (n)
    = 
    \sum_i^{N}
    \sum_h^{N}
    \hat{r}_{h \to i}\, \hat{L}_{h \to i} (n)
    + \hat{b}\, I_s (n)
    \, ,
    \label{eq:ART output}
\end{equation}
which also includes the direct-incidence operator, $\hat{b}$.
Together, Eqs.~\eqref{eq:ART as FDN},~\eqref{eq:ART injection},~and~\eqref{eq:ART output} form the complete \ac{ART} acoustic model.

\subsection{Delay-based modeling}

Delay-based acoustic models, introduced in~\cite{schroeder_colorless, schroeder_natural}, rely on inter-connected networks of delay lines to generate an output with an echo-rich structure.
They circumvent the need for convolution with \acp{RIR}, which considerably reduces computational complexity under most conditions of practical significance.
Among these models are \acf{FDN} structures~\cite{jot_fdn}, which revolve around a set of delay lines ($d_i$) and a feedback matrix ($\bm{A}$).
In an \ac{FDN} with $M$ delay lines, the system output $p_r$ is
\begin{align}
    p_r (n)
    = & \ 
    \sum_i^{M}
    r_i\, p_i (n)
    + b\, p_s (n)
    \, ,
    \label{eq:FDN output}
    \\
    p_i (n)
    = & \ 
    \sum_h^{M}
    A_{i, h}\, p_h (n - d_i)
    + s_i\, p_s (n - d_i)
    \, ,
    \label{eq:FDN state}
\end{align}
where $p_s$ is the system input, while $s_i$, $r_i$, and $b$ are input and output weights (i.e.\@ scalars, not operators).
If ${d_i = 1 \, \forall i}$, Eq.~\eqref{eq:FDN output},~\eqref{eq:FDN state} correspond to the measurement and state equations of a state-space model, respectively.
In other words, an \ac{FDN} corresponds to a generalized version of a state-space model with non-unit delays~\cite{rocchesso_fdn}.

An important aspect of \ac{FDN} design is ensuring the system's stability, meaning that its output is bounded for any bounded input.
All of the works described in the following employ so-called \emph{unilossless} feedback matrices, $\bm{A}$, which guarantee stability for arbitrary (time-invariant) delay line lengths~\cite{unilossless}.

\begin{figure}[t]
    \begin{adjustbox}{max width=0.48\textwidth}
    \input{figures/fdn_blocks_wide}
    \end{adjustbox}
    \caption{Block diagram of the delay-based model structure used in this paper. Individual blocks can be either simple time delays and scaling coefficients, or more elaborate \ac{FIR} filters. Note that this structure does not exactly match that of \acp{FDN}, because of the position of the delay operators in the loop and the presence of filters associated to each input and output.}
    \label{fig:fdn blocks}
\end{figure}
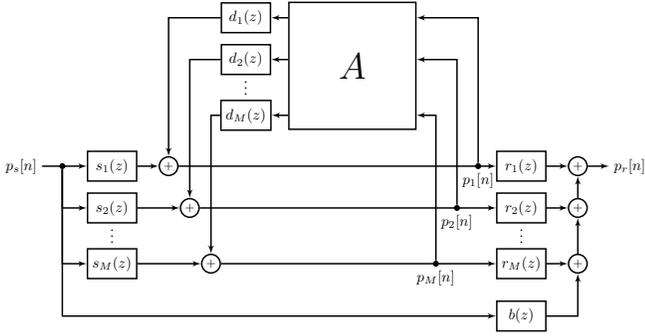

Most delay-based methods lack explicit physical parameters, and are instead calibrated based on the desired reverberation properties~\cite{fifty_years}.
Recent work, discussed in the following, introduced \acl{GA} concepts in the design of delay-based models.
Note that the network structures of the discussed models, including the one proposed in this paper, are similar but not identical to the standard \ac{FDN} structure.
This is due in part to the presence of additional filters, and in part to the position of the delay operators in the feedback loop.
In particular, the structure we employ (illustrated in \figurename~\ref{fig:fdn blocks}) is designed to plainly display its links with \ac{ART}.

\subsubsection{\aclp{ARN}}

\acfp{ARN}~\cite{arn} leverage the similarity of the \ac{ART} definition (Eq.~\eqref{eq:ART as FDN},~\eqref{eq:ART output}) to the \ac{FDN} structure (Eq.~\eqref{eq:FDN output},~\eqref{eq:FDN state}), designing a delay-based model using a surface discretization akin to \ac{ART}.
The main difference between the two is that the \ac{ART} equations are intensity-based quantities (i.e.\@ squared pressure), while delay-based models typically propagate pressure quantities.
Scaling coefficients for air and surface materials are square-rooted to match this change in signal domain.
As a result, the feedback signals $p_{h \to i}$ traveling over the delay lines
can be interpreted as root-radiance quantities:
\begin{align}
    p_{h \to i} (n)
    = & \ 
    \pm \sqrt{\hat{L}_{h \to i} (n)}
    \, .
    \label{eq:root radiance}
\end{align}
Note that phase information, i.e.\@ sign, is not preserved; this aspect is shared by all models which employ squared-pressure signals.
Following the definition of \ac{ART}, delay line lengths are based on the delay components $\bm{D}$, one for each path (pair of patches).
The feedback matrix is only loosely based on $\bm{\hat{S}}$, and does not offer control of scattering coefficients -- this will be discussed in more detail at the end of this section.
Air and material absorption can be modeled by pairing each delay line with a filter, possibly frequency-dependent.
The inputs of delay lines are weighted based on the injection operators (Eq.~\eqref{eq:ART injection}), and the outputs are likewise weighted for detection (Eq.~\eqref{eq:ART output}).

\subsubsection{\acs{FDN}-\acs{ART}}

\begin{figure*}[t]
    \centering
    \subfloat[Full matrix $\bm{\hat{S}}$]{%
        \begin{adjustbox}{max width=0.4\textwidth}%
        \input{figures/matrix_full}%
        \end{adjustbox}%
        \label{fig:ART matrix full}%
    }
    \hfil
    \subfloat[Matrix block $\bm{\hat{S}}_1$.]{%
        \begin{adjustbox}{max width=0.35\textwidth}%
        \input{figures/matrix_block}%
        \end{adjustbox}%
        \label{fig:ART matrix block}%
    }
    \caption{$\bm{\hat{S}}$ matrix for a ``hallway'' room of size \qtyproduct{2 x 2 x 6}{\metre}, reflection coefficient $0.9$, and scattering coefficient $0.25$.
    Note that values of $0$ are drawn in white. The elements contoured by red dashed lines in~\ref{fig:ART matrix full} make up the block $\bm{\hat{S}}_1$ shown in~\ref{fig:ART matrix block}, where 1 is the floor patch index.
    }
    \label{fig:ART matrix}
\end{figure*}
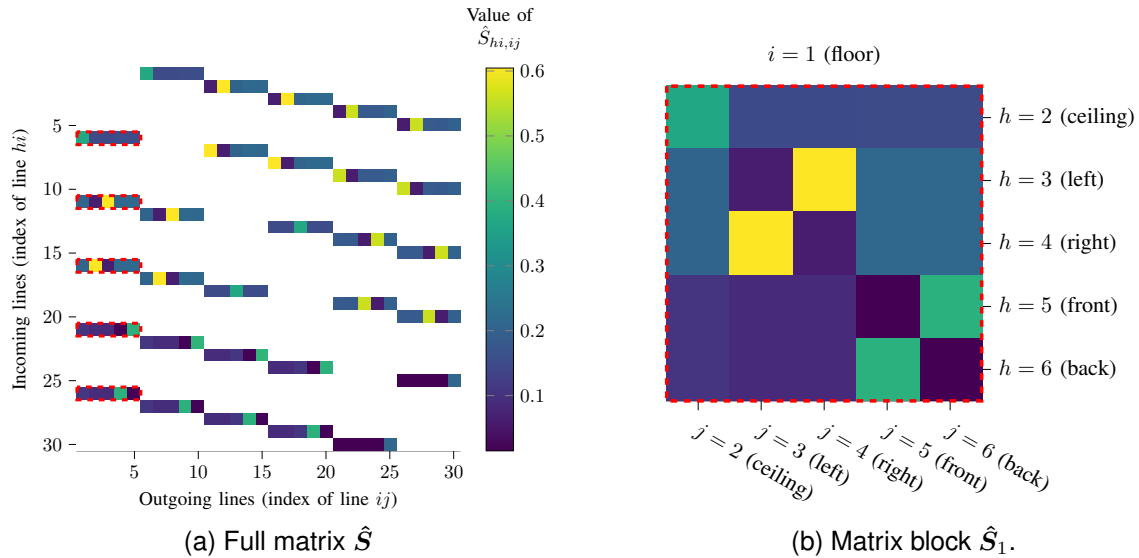

The \acs{FDN}-\acs{ART} method~\cite{fdn_art} also starts from a surface discretization, but rather than trying to match the delay-based structure of \ac{ART}, it generates a small number of delay lines based on the most commonly occurring values in $\bm{D}$.
The process aims to detect the environment's salient modes, by selecting delay line lengths related to the major resonance frequencies.
The details of the feedback matrix design are not relevant in the context of this paper, as the resulting matrices do not preserve spatial information (lines don't correspond to physical paths).
As in \ac{ARN}, wall and air absorption are modeled by filters associated to each delay line.

The network structure is designed in such a way that the feedback loop only generates reflections from the second reflection order onwards, while first order reflections are modeled by additional delay lines bypassing the loop.
For this reason, first order reflections may be modeled by any \ac{GA} method of choice, depending on the desired accuracy of the early impulse response.

\subsection{ARN matrix design and sparseness of ART matrix}

Before discussing the design of \ac{ARN} recursion matrices, $\bm{A}$, it will be useful to illustrate some properties of \ac{ART} recursion matrices, $\bm{\hat{S}}$.
Aside from providing insight on the functioning and design of $\bm{A}$, this will show how the cost of operations on these matrices can be reduced from $N^4$ to $N^3$ or less.

We will examine the structure of $\bm{\hat{S}}$ with the help of an example matrix (shown in \figurename~\ref{fig:ART matrix}) related to a cuboid room of size \qtyproduct{2 x 2 x 6}{\metre}, reflection coefficient $0.9$, and scattering coefficient $0.25$.
The room is divided in 6 surface patches, one for each wall.
Note that this is an extremely coarse discretization, for illustrative purposes; much smaller patches are generally used in \ac{ART}.
The elements of $\bm{\hat{S}}$ contoured by red dashed lines in \figurename~\ref{fig:ART matrix full} form the sub-matrix $\bm{\hat{S}}_1$ shown in \figurename~\ref{fig:ART matrix block}.
The contoured rows are those related to lines which end at patch 1 (the floor), and the contoured columns are those related to lines which start from the same patch.
Each block $\bm{\hat{S}}_i$ defined this way characterizes the scattering behaviour of the corresponding patch $i$.
With this in mind, the reason why $\bm{\hat{S}}$ is so highly sparse (most elements are 0) is that any line terminating at a given patch $i$ only provides contributions to the lines which depart from the same patch, i.e.
\begin{align}
    \hat{S}_{h \to i, j \to k}
    = 0
    \quad \forall i \ne j
    \ ,
    \label{eq:S blocks 1}
\end{align}
because a patch can only scatter energy to and from the paths connected to it.
This observation leads to a considerable reduction in the computational cost of designing and employing matrices based on $\bm{\hat{S}}$, as we will discuss in the following.

It is possible for even more elements of $\bm{\hat{S}}$ to be 0.
If there is no direct visibility between a pair of patches $h, i$, two rows and two columns will be entirely 0 -- those corresponding to the two paths ${h \to i}$, ${i \to h}$, which relate to one row and one column each.
Of course, paths without visibility carry no energy in either direction, and the redundant delay lines should be removed from the network along with the corresponding matrix entries.
The same applies to paths connecting each patch to itself, or to any other patch located on the same flat surface.
In \figurename~\ref{fig:ART matrix}, self-interaction paths have already been removed.

With these properties in mind, we can now clearly see how the complexity of matrix operations can be reduced from $N^4$ to $N^3$ or less.
With $N$ surface patches, there are up to $N^2$ delay lines (patch pairs), therefore up to $N^4$ elements in $\bm{\hat{S}}$.
Due to Eq.~\eqref{eq:S blocks 1}, nonzero elements are $N^3$ in total ($N$ blocks of size ${N \times N}$ each).
Lack of visibility may further reduce the size of each block.
Awareness of the null elements does not only reduce the computational complexity of running the recursive system, but also the complexity of the matrix design.

The original \ac{ARN}~\cite{arn} makes use of a permuted Householder matrix for each block $\bm{A}_i$ of the feedback matrix $\bm{A}$.
The first step in this design is the evaluation of a permutation matrix, $\bm{P}_i$.
The evaluation, detailed in the next paragraph, is such that the nonzero elements of $\bm{P}_i$ correspond to pairs of delay lines which form specular reflections.
Then, in each block $\bm{A}_i$, the elements chosen as ``specular'' are set to ${(2-M_i) / M_i}$, where $M_i$ is the size of the block $\bm{A}_i$, and the remaining elements\footnote{Note that paths without visibility were removed: all elements in each block $\bm{A}_i$ correspond to patches which are visible from patch $i$.} are set to ${2 / M_i}$.
As a result, $\bm{A}_i$ is a permuted Householder matrix, and the elements related to specular reflections carry more energy than others.

\begin{figure}[t]
    \centering
    \begin{adjustbox}{max width=0.3\textwidth}
    \input{figures/permutation_fail}
    \end{adjustbox}
    \caption{A possible situation where ``naive'' pairing of specular paths fails to produce a permutation matrix. Given patch $i$ as a reflector, it is clear that the incoming direction ${h \to i}$ holds the most specularly reflected energy for both outgoing directions ${i \to j}$, ${i \to k}$. For visual clarity, very few sample points are shown on the areas $A_i$, $A_j$, $A_k$; evaluating Eq.~\eqref{eq:discretized kernel} with fine discretization does not prevent the effect.}
    \label{fig:permutation fail}
\end{figure}
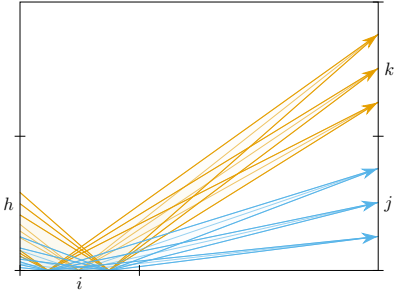

It should be noted that the creation of the ``specular'' permutation matrix $\bm{P}_i$ is not trivial.
Bai \textit{et al}.~\cite{arn} select for each column in $\bm{\hat{S}}$ the path which receives the most specularly reflected energy.
However, there are several instances where this strategy does not yield an actual permutation matrix, and it is unclear how these cases are handled in~\cite{arn}.
An example of a problematic situation is illustrated in \figurename~\ref{fig:permutation fail}.
In the example, two different columns (corresponding to lines ${i \to j}$ and ${i \to k}$) would select the same row (corresponding to line ${h \to i}$) as the direction of most specularly reflected energy.
With more than one non-zero element on a single row, the result is not a permutation matrix.
This is, in fact, the case for all but the most trivial geometry discretizations.

The strategy used in the remainder of this paper to obtain such permutation matrices consists of iteratively selecting the maximum value in $\bm{\hat{S}}$, setting the corresponding element to 1 in the matrix, and ignoring the selected row/column pair in subsequent iterations.
The resulting matrix is guaranteed to be a permutation matrix, and the conflicts illustrated above are resolved by prioritizing higher energy transfers.

In the next section, we will discuss the limitations of the permuted Householder matrix design, and propose two methods to overcome them.

%% file: figures/points_on_surfs.tex
\begin{tikzpicture}[3d view]
    \coordinate (point_x) at (4.0, 2.5, 1.7);
    \coordinate (point_x_prime) at (0.8, 0.8, 0.0);
    \coordinate (point_x_second) at (1.5, 5.0, 0.5);
    
    \fill[gray, opacity=0.5] (tpp cs:x=0,y=5,z=0)
    -- (tpp cs:x=0,y=5,z=1)
    -- (tpp cs:x=2,y=5,z=1)
    -- (tpp cs:x=2,y=5,z=0) -- cycle;

	\draw[thin] (0,5,0) -- (0,5,1);
	\draw[thin] (0,5,1) -- (2,5,1);
	\draw[thin] (2,5,1) -- (2,5,0);
	\draw[thin] (2,5,0) -- (0,5,0);

    \fill (point_x_second) circle (1.5pt);
    \fill[black,font=\footnotesize] (point_x_second) node [above] {$x''$};
    
    \fill[gray, opacity=0.5] (tpp cs:x=0,y=0,z=0)
    -- (tpp cs:x=0,y=2,z=0)
    -- (tpp cs:x=2,y=2,z=0)
    -- (tpp cs:x=2,y=0,z=0) -- cycle;

	\draw[thin] (0,0,0) -- (2,0,0);
	\draw[thin] (2,0,0) -- (2,2,0);
	\draw[thin] (2,2,0) -- (0,2,0);
	\draw[thin] (0,2,0) -- (0,0,0);

    \fill (point_x_prime) circle (1.5pt);
    \fill[black,font=\footnotesize] (point_x_prime) node [left] {$x'$};
    
	\draw[-stealth] (point_x_prime) -- +(0,0,1);
    \fill[black,font=\footnotesize] (point_x_prime)+(0.1,0,0.9) node [above] {$n(x')$};
    
    \fill[gray, opacity=0.5]  (tpp cs:x=4,y=1,z=1)
    -- (tpp cs:x=4,y=1,z=2)
    -- (tpp cs:x=4,y=3,z=2)
    -- (tpp cs:x=4,y=3,z=1) -- cycle;

	\draw[thin] (4,1,1) -- (4,1,2);
	\draw[thin] (4,1,2) -- (4,3,2);
	\draw[thin] (4,3,2) -- (4,3,1);
	\draw[thin] (4,3,1) -- (4,1,1);

    \fill (point_x) circle (1.5pt);
    \fill[black,font=\footnotesize] (point_x) node [right] {$x$};
    
	\draw[-stealth] (point_x) -- +(-1,0,0);
    \fill[black,font=\footnotesize] (point_x)+(-0.7,0,0) node [above] {$n(x)$};
    
	\draw[dashed, opacity=0.5] (point_x) -- (point_x_prime);
	\draw[dashed, opacity=0.5] (point_x_prime) -- (point_x_second);

    \draw[-stealth] (point_x) -- ($(point_x)!1cm!(point_x_prime)$);
    \draw[-stealth] (point_x_prime) -- ($(point_x_prime)!1cm!(point_x)$);
    \draw[-stealth] (point_x_prime) -- ($(point_x_prime)!1cm!(point_x_second)$);
    
    \fill[black,font=\footnotesize] ($(point_x)!0.7cm!(point_x_prime)$) node [left] {$v(x|x')$};
    \fill[black,font=\footnotesize] ($(point_x_prime)!0.7cm!(point_x)$) node [right] {$v(x'|x)$};
    \fill[black,font=\footnotesize] ($(point_x_prime)!0.6cm!(point_x_second)$) node [left] {$v(x'|x'')$};

\end{tikzpicture}

%% file: figures/fdn_blocks_wide.tex
\begin{tikzpicture}[auto,>=latex']
    \tikzstyle{block} = [draw, shape=rectangle, minimum height=2em, minimum width=3.3em, node distance=2cm, line width=1pt]
    
    \tikzstyle{matrix} = [draw, shape=rectangle, minimum height=3cm, minimum width=3cm, node distance=2cm, line width=1pt]
    
    \tikzstyle{sum} = [draw, shape=circle, node distance=1.5cm, line width=1pt, minimum width=1.25em]
    
    \tikzstyle{branch}=[fill,shape=circle,minimum size=4pt,inner sep=0pt]
    
    \node at (0,0) (input) {$p_s[n]$};
    \node [branch, right = 0.4cm of input] (branch_in) {};
    
    \node [block, right = 0.5cm of branch_in] (b1) {$s_1(z)$};
    \node [sum, right = 0.5cm of b1] (sum1) {};
    \node at (sum1) (plus) {{\footnotesize$+$}};
    \node [branch, right = 7.0cm of sum1] (branch1) {};
    \node [block, right = 7.5cm of sum1] (c1) {$r_1(z)$};
    \node [sum, right = 0.5cm of c1] (sum_out1) {};
    \node at (sum_out1) (plus) {{\footnotesize$+$}};
    
    \node [block, below = 0.25cm of b1] (b2) {$s_2(z)$};
    \node [sum, right = 1.0cm of b2] (sum2) {};
    \node at (sum2) (plus) {{\footnotesize$+$}};
    \node [branch, right = 6.0cm of sum2] (branch2) {};
    \node [block, below = 0.25cm of c1] (c2) {$r_2(z)$};
    \node [sum, right = 0.5cm of c2] (sum_out2) {};
    \node at (sum_out2) (plus) {{\footnotesize$+$}};
    
    \node [below = -0.2cm of b2] (bd) {$\vdots$};
    \node [below = -0.2cm of c2] (cd) {$\vdots$};
    
    \node [block, below = 0.0cm of bd] (bM) {$s_M(z)$};
    \node [sum, right = 1.5cm of bM] (sumM) {};
    \node at (sumM) (plus) {{\footnotesize$+$}};
    \node [branch, right = 5.0cm of sumM] (branchM) {};
    \node [block, below = 0.0cm of cd] (cM) {$r_M(z)$};
    \node [sum, right = 0.5cm of cM] (sum_outM) {};
    \node at (sum_outM) (plus) {{\footnotesize$+$}};
    
    \node [matrix, above left = 0.8cm and 1.4cm of branch1] (A) {\Huge$A$};
    
    \node [block, above left = -0.75cm and 0.4cm of A] (s1) {$d_1(z)$};
    \node [block, below = 0.25cm of s1] (s2) {$d_2(z)$};
    \node [below = -0.2cm of s2] (sd) {$\vdots$};
    \node [block, below = 0.0cm of sd] (sM) {$d_M(z)$};
    
    \node [block, below = 0.5cm of cM] (d) {$b(z)$};
    \node [right = 0.5cm of sum_out1] (output) {$p_r[n]$};
    
    \node [below = -0.05cm of branch1] (l1) {$p_1[n]$};
    \node [below = -0.05cm of branch2] (l2) {$p_2[n]$};
    \node [below = -0.05cm of branchM] (lM) {$p_M[n]$};
    
    \begin{scope}[line width=1pt]
         \draw[->] (input) -- (b1);
         \draw[->] (branch_in) |- (d);
         \draw[->] (branch_in) |- (b2);
         \draw[->] (branch_in) |- (bM);
         
         \draw[->] (b1) -- (sum1);
         \draw[->] (b2) -- (sum2);
         \draw[->] (bM) -- (sumM);
         
         \draw[->] (s1.east -| A.west) -- (s1.east);
         \draw[->] (s1) -| (sum1);
         \draw[->] (s2.east -| A.west) -- (s2.east);
         \draw[->] (s2) -| (sum2);
         \draw[->] (sM.east -| A.west) -- (sM.east);
         \draw[->] (sM) -| (sumM);
         
         \draw[->] (branch1) |- (s1.east -| A.east);
         \draw[->] (branch2) |- (s2.east -| A.east);
         \draw[->] (branchM) |- (sM.east -| A.east);
         
         \draw[->] (sum1) -- (c1);
         \draw[->] (sum2) -- (c2);
         \draw[->] (sumM) -- (cM);
         
         \draw[->] (c1) -- (sum_out1);
         \draw[->] (c2) -- (sum_out2);
         \draw[->] (cM) -- (sum_outM);
         \draw[->] (d) -| (sum_outM);
         \draw[->] (sum_outM) -- (sum_out2);
         \draw[->] (sum_out2) -- (sum_out1);
         \draw[->] (sum_out1) -- (output);
    \end{scope}
\end{tikzpicture}

%% file: figures/matrix_full.tex
\begin{tikzpicture}

\begin{axis}[
clip=false,
colorbar,
colorbar style={align=center, title={Value of\\$\hat{S}_{hi, ij}$}},
colormap/viridis,
width=7cm,
height=7cm,
scale only axis,
point meta max=0.604565185143766,
point meta min=0.014870717985504,
tick align=outside,
tick pos=left,
x grid style={darkgray},
xlabel={Outgoing lines (index of line $ij$)},
xmin=0.5, xmax=30.5,
xtick style={color=black},
y dir=reverse,
y grid style={darkgray},
ylabel={Incoming lines (index of line $hi$)},
ymin=0.5, ymax=30.5,
ytick style={color=black}
]
\addplot graphics [xmin=0.5, xmax=30.5, ymin=30.5, ymax=0.5] {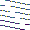};

\draw [ultra thick, dashed, red] (0.5, 5.5) rectangle (5.5, 6.5);
\draw [ultra thick, dashed, red] (0.5, 10.5) rectangle (5.5, 11.5);
\draw [ultra thick, dashed, red] (0.5, 15.5) rectangle (5.5, 16.5);
\draw [ultra thick, dashed, red] (0.5, 20.5) rectangle (5.5, 21.5);
\draw [ultra thick, dashed, red] (0.5, 25.5) rectangle (5.5, 26.5);

\end{axis}

\end{tikzpicture}

%% file: figures/matrix_block.tex
\begin{tikzpicture}

\begin{axis}[
clip=false,
width=5cm,
height=5cm,
scale only axis,
tick align=outside,
xtick pos=bottom,
ytick pos=right,
title={$i=1$ (floor)},
x grid style={darkgray},
xmin=0.5, xmax=5.5,
xtick style={color=black},
y dir=reverse,
y grid style={darkgray},
ymin=0.5, ymax=5.5,
ytick style={color=black},
ytick={1,2,3,4,5},
yticklabels={${h=2}$ (ceiling),
             ${h=3}$ (left),
             ${h=4}$ (right),
             ${h=5}$ (front),
             ${h=6}$ (back),},
xtick={1,2,3,4,5},
xticklabels={${j=2}$ (ceiling),
             ${j=3}$ (left),
             ${j=4}$ (right),
             ${j=5}$ (front),
             ${j=6}$ (back)},
x tick label style={rotate=-30, anchor=north west},
]
\addplot graphics [xmin=0.5, xmax=5.5, ymin=5.5, ymax=0.5]{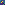};

\draw [ultra thick, dashed, red] (0.5, 0.5) rectangle (5.5, 5.5);

\end{axis}

\end{tikzpicture}

%% file: figures/permutation_fail.tex
\begin{tikzpicture} [x=1, y=1, rotate=90]
    \definecolor{color1}{RGB}{230, 159, 0}; 
    \definecolor{color2}{RGB}{86, 180, 233}; 
    \definecolor{color3}{RGB}{0, 158, 115}; 
    \definecolor{color4}{RGB}{240, 228, 66}; 
    \definecolor{color5}{RGB}{0, 114, 178}; 
    \definecolor{color6}{RGB}{213, 94, 0}; 
    \definecolor{color7}{RGB}{204, 121, 167}; 
    
    \def \roomX {150};
    \def \roomY {200};

    \def \spacingI {17};
    \def \spacingJ {19};

    \def \rayOpacity {0.6};
    \def \beamOpacity {0.02};
    
    \coordinate (Corner_NW) at (0, \roomY);
    \coordinate (Corner_NE) at (\roomX, \roomY);
    \coordinate (Corner_SE) at (\roomX, 0);
    \coordinate (Corner_SW) at (0, 0);
    
    \coordinate (Patch_h_west) at (0, \roomY);
    \coordinate (Patch_h_east) at (\roomX/2, \roomY);
    \coordinate (Patch_i_south) at (0, 2*\roomY/3);
    \coordinate (Patch_i_north) at (0, \roomY);
    \coordinate (Patch_j_west) at (0, 0);
    \coordinate (Patch_j_east) at (\roomX/2, 0);
    \coordinate (Patch_k_west) at (\roomX/2, 0);
    \coordinate (Patch_k_east) at (\roomX, 0);

    \coordinate (Patch_h_center) at ($(Patch_h_west)!0.5!(Patch_h_east)$);
    \coordinate (Patch_i_center) at ($(Patch_i_south)!0.5!(Patch_i_north)$);
    \coordinate (Patch_j_center) at ($(Patch_j_west)!0.5!(Patch_j_east)$);
    \coordinate (Patch_k_center) at ($(Patch_k_west)!0.5!(Patch_k_east)$);

    \foreach \a in {0,\spacingJ,...,75}{
        \ifnum\a=0{
        }\else{
            \ifnum\a=75{
            }\else{
                \edef \minrayX {-1};
                \edef \maxrayX {-1};
                \edef \minrayY {-1};
                \edef \maxrayY {-1};
                
                \coordinate (k_point) at ([xshift=\a]Patch_k_west);
                
                \foreach \b in {0,\spacingI,...,65}{
                    \ifnum\b=0{
                    }\else{
                        \ifnum\b=65{
                        }\else{
                            \coordinate (i_point) at ([yshift=\b]Patch_i_south);
                            \path [overlay, name path=extended] (k_point) -- ($(k_point)!2!(i_point)$);
                            \path [overlay, name path=image_h] (Corner_NW) -- ([xshift=-\roomX]Corner_NW);
                            \path [overlay, name intersections={of=extended and image_h, by=inters}];
                            \newdimen\xOne;
                            \newdimen\xTwo;
                            \pgfextractx{\xOne}{\pgfpointanchor{inters}{center}};
                            \pgfextractx{\xTwo}{\pgfpointanchor{Patch_h_east}{center}};
                            \ifthenelse{-\xOne < \xTwo}{
                                \coordinate (h_point) at ([xshift=-\xOne]Patch_h_west);
                                \ifthenelse{\minrayY < 0}{
                                    \pgfmathparse {\b};
                                    \xdef\minrayY {\pgfmathresult};
                                    \pgfmathparse {-\xOne};
                                    \xdef\minrayX {\pgfmathresult};
                                }{};
                                \ifthenelse{\maxrayY < \b}{
                                    \pgfmathparse {\b};
                                    \xdef\maxrayY {\pgfmathresult};
                                    \pgfmathparse {-\xOne};
                                    \xdef\maxrayX {\pgfmathresult};
                                }{};
                                
                                \draw [opacity=\rayOpacity, draw=color1] (k_point) -- (i_point);
                                \draw [opacity=\rayOpacity, draw=color1] (i_point) -- (h_point);
                            }{};
                        }\fi;
                    }\fi;
                };
                \coordinate (i_point_1) at ([yshift=\minrayY]Patch_i_south);
                \coordinate (i_point_2) at ([yshift=\maxrayY]Patch_i_south);
                \coordinate (h_point_1) at ([xshift=\minrayX]Patch_h_west);
                \coordinate (h_point_2) at ([xshift=\maxrayX]Patch_h_west);
                
                \draw [fill=color1, opacity=\beamOpacity, draw=none] (k_point) -- (i_point_1) -- (i_point_2) -- cycle;
                \draw [fill=color1, opacity=\beamOpacity, draw=none] (i_point_1) -- (i_point_2) -- (h_point_2) -- (h_point_1) -- cycle;
    
                \draw [arrows={-Stealth[harpoon, length=3mm]}, color=color1] (i_point_1) -- (k_point);
                \draw [arrows={-Stealth[harpoon, swap, length=3mm]}, color=color1] (i_point_2) -- (k_point);
                \draw [color=color1] (h_point_2) -- (i_point_2);
                \draw [color=color1] (h_point_1) -- (i_point_1);
            }\fi;
        }\fi;
    };

    \foreach \a in {0,\spacingJ,...,75}{
        \ifnum\a=0{
        }\else{
            \ifnum\a=75{
            }\else{
                \edef \minrayX {-1};
                \edef \maxrayX {-1};
                \edef \minrayY {-1};
                \edef \maxrayY {-1};
                
                \coordinate (j_point) at ([xshift=\a]Patch_j_west);
                
                \foreach \b in {0,\spacingI,...,65}{
                    \ifnum\b=0{
                    }\else{
                        \ifnum\b=65{
                        }\else{
                            \coordinate (i_point) at ([yshift=\b]Patch_i_south);
                            \path [overlay, name path=extended] (j_point) -- ($(j_point)!2!(i_point)$);
                            \path [overlay, name path=image_h] (Corner_NW) -- ([xshift=-\roomX]Corner_NW);
                            \path [overlay, name intersections={of=extended and image_h, by=inters}];
                            \newdimen\xOne;
                            \newdimen\xTwo;
                            \pgfextractx{\xOne}{\pgfpointanchor{inters}{center}};
                            \pgfextractx{\xTwo}{\pgfpointanchor{Patch_h_east}{center}};
                            \ifthenelse{-\xOne < \xTwo}{
                                \coordinate (h_point) at ([xshift=-\xOne]Patch_h_west);
                                \ifthenelse{\minrayY < 0}{
                                    \pgfmathparse {\b};
                                    \xdef\minrayY {\pgfmathresult};
                                    \pgfmathparse {-\xOne};
                                    \xdef\minrayX {\pgfmathresult};
                                }{};
                                \ifthenelse{\maxrayY < \b}{
                                    \pgfmathparse {\b};
                                    \xdef\maxrayY {\pgfmathresult};
                                    \pgfmathparse {-\xOne};
                                    \xdef\maxrayX {\pgfmathresult};
                                }{};
                                
                                \draw [opacity=\rayOpacity, draw=color2] (j_point) -- (i_point);
                                \draw [opacity=\rayOpacity, draw=color2] (i_point) -- (h_point);
                            }{};
                        }\fi;
                    }\fi;
                };
                \coordinate (i_point_1) at ([yshift=\minrayY]Patch_i_south);
                \coordinate (i_point_2) at ([yshift=\maxrayY]Patch_i_south);
                \coordinate (h_point_1) at ([xshift=\minrayX]Patch_h_west);
                \coordinate (h_point_2) at ([xshift=\maxrayX]Patch_h_west);
                
                \draw [fill=color2, opacity=\beamOpacity, draw=none] (j_point) -- (i_point_1) -- (i_point_2) -- cycle;
                \draw [fill=color2, opacity=\beamOpacity, draw=none] (i_point_1) -- (i_point_2) -- (h_point_2) -- (h_point_1) -- cycle;
    
                \draw [arrows={-Stealth[harpoon, length=3mm]}, color=color2] (i_point_1) -- (j_point);
                \draw [arrows={-Stealth[harpoon, swap, length=3mm]}, color=color2] (i_point_2) -- (j_point);
                \draw [color=color2] (h_point_2) -- (i_point_2);
                \draw [color=color2] (h_point_1) -- (i_point_1);
            }\fi;
        }\fi;
    };

    \draw (Corner_NW) -- (Corner_NE) -- (Corner_SE) -- (Corner_SW) -- cycle;
    
    \draw ([yshift=3]Patch_h_west) -- ([yshift=-3]Patch_h_west);
    \draw ([yshift=3]Patch_h_east) -- ([yshift=-3]Patch_h_east);
    \draw ([xshift=3]Patch_i_south) -- ([xshift=-3]Patch_i_south);
    \draw ([xshift=3]Patch_i_north) -- ([xshift=-3]Patch_i_north);
    \draw ([yshift=3]Patch_j_west) -- ([yshift=-3]Patch_j_west);
    \draw ([yshift=3]Patch_j_east) -- ([yshift=-3]Patch_j_east);
    \draw ([yshift=3]Patch_k_east) -- ([yshift=-3]Patch_k_east);
    
    \draw (Patch_h_center) node [anchor=east] {$h$};
    \draw (Patch_i_center) node [anchor=north] {$i$};
    \draw (Patch_j_center) node [anchor=west] {$j$};
    \draw (Patch_k_center) node [anchor=west] {$k$};
\end{tikzpicture}

%% file: sections/proposed.tex
\section{Proposed model}
\label{sec:proposed}

The extensions proposed in this paper are based on the \ac{ARN} interpretation of feedback signals as root-radiance quantities (see Eq.~(\ref{eq:root radiance})).
This interpretation can be leveraged to obtain higher accuracy and flexibility in the recursive component of reverberation, as well as in the early reflections.
The first extension, presented in Section~\ref{sec:proposed-matrices}, is focused on the design of the feedback matrix $\bm{A}$, with the objective of obtaining more accurate reverberation times and preservation of spatial information (i.e.\@ energy flowing over each path in space, at each point in time).
The second extension, presented in Section~\ref{sec:proposed-injection}, is focused on the design of the input operators $s_{ij}$, with the objective of obtaining more accurate early reflections.
Section~\ref{sec:proposed-parameters} presents the design of the remaining system components (recursion, detection, and bypassing operators).

\subsection{Physically-driven feedback matrix designs}
\label{sec:proposed-matrices}

The Householder matrix design employed in~\cite{arn} has two drawbacks.
The first is that it offers no control of scattering coefficients.
Instead, the ratio of specular-to-diffuse energy is controlled entirely by the size of matrix blocks, $M_i$, which, in turn, is set by the surface discretization resolution.
The second drawback, also related to the block size $M_i$, is that the Householder matrices tend towards simple identity matrices as $M_i$ increases.
Note that high surface discretization resolution is required for good modeling accuracy of \ac{ART}: attempting to increase the spatial resolution of the model induces the aforementioned artefacts.
This has adverse effects on the reverberation times and echo density produced by the \ac{ARN}, as will be shown in Section~\ref{sec:results}.

The objective of the extensions proposed in this section is to design the feedback matrix $\bm{A}$  such that it achieves a closer match to the \ac{ART} matrix $\bm{\hat{S}}$, all the while preserving stability.
To enable the use of arbitrary scattering coefficients (and in theory, arbitrary \acp{BRDF}), we propose adapting the \emph{closest sign-agnostic matrix} optimization process introduced by Schlecht and Habets~\cite{sign_agnostic}.
The process is intended to generate the unilossless feedback matrix $\bm{A}$ which most closely matches a given target matrix $\bm{\Tilde{A}}$, with a focus on the absolute values of entries, which govern energy distribution.
The sign-agnostic optimization is restricted to target matrices such that their Hadamard (i.e.\@ element-wise) square $\bm{\Tilde{A}} \circ \bm{\Tilde{A}}$ is doubly-stochastic (i.e.\@ each row sums to 1, as do the columns)\footnote{Note that if there exists a non-singular diagonal matrix $\bm{E}$ such that ${\bm{B} = \bm{E} \bm{\Tilde{A}} \bm{E}^{-1}}$ is doubly-stochastic, using $\bm{B}$ in place of $\bm{\Tilde{A}}$ and compensating for $\bm{E}$ in the input-output parameters results in an equivalent system (see Theorem~3~in~\cite{unilossless}).}.
Unfortunately, although the columns of $\bm{\hat{S}}$ sum to 1 (see Eq.~(\ref{eq:R energy conservation})), the same is not necessarily true of its rows.

In order to apply the closest sign-agnostic optimization here, it is necessary to first find the closest doubly-stochastic matrix to $\bm{\hat{S}}$.
Before proposing two strategies for this problem, it is useful to note that because of the property in Eq.~(\ref{eq:S blocks 1}), $\bm{\hat{S}}$ is doubly stochastic if and only if each block $\bm{\hat{S}}_i$ is doubly stochastic.
This is because all nonzero elements lying on a given row $hi$ in $\bm{\hat{S}}$ define one row of one block $\bm{\hat{S}}_i$, and all nonzero elements lying on a given column $ij$ define one column of one block $\bm{\hat{S}}_i$  (this can be verified visually in \figurename~\ref{fig:ART matrix full}).
As a result, either one of the proposed approaches can be applied to individual blocks $\bm{\hat{S}}_i$ to greatly reduce operations.

Two separate design strategies are presented here and compared in Section~\ref{sec:results}.
One strategy results in a higher accuracy but can fail under certain conditions, while the other is more approximate but prevents such failure conditions.

\subsubsection{Sinkhorn-Knopp matrix design}
\label{sec:proposed-matrices-sinkhorn}

The first approach uses the Sinkhorn-Knopp algorithm~\cite{sinkhorn_knopp}, which provides the closest doubly-stochastic matrix to a target.
More precisely, given the target $\bm{\hat{S}}_i$, Sinkhorn-Knopp provides the diagonal matrices $\bm{E}_1, \bm{E}_2$ such that ${\bm{\Tilde{A}}_i \circ \bm{\Tilde{A}}_i = \bm{E}_1 \bm{\hat{S}}_i \bm{E}_2}$ is doubly-stochastic.

A necessary and sufficient condition for the Sinkhorn-Knopp algorithm to converge is that the target matrix has total support~\cite{sinkhorn_knopp, sinkhorn_convergence}.
There are some cases where this is not the case for $\bm{\hat{S}}_i$, and Sinkhorn-Knopp is not applicable.
For example, if there is no diffuse component to the \ac{BRDF} of patch $i$, energy traversing a path $hi$ will be reflected towards a few -- but not all -- patches visible from $i$.
As a result, different rows in $\bm{\hat{S}}_i$ may have different numbers of nonzero elements, and the block may not have total support, leading to divergence (elements in $\bm{E}_1$, $\bm{E}_2$ tend to infinity).

Note how this strategy allows modeling of arbitrary scattering coefficients as well as arbitrary \acp{BRDF}.

\subsubsection{Uniform matrix design}
\label{sec:proposed-matrices-uniform}

The second approach uses a simplified scattering model.
Similarly to the Householder matrices used in~\cite{arn}, the starting point is a permutation matrix $\bm{P}_i$ correlating the pairs of paths that carry the most specular energy.
Then, for each block $\bm{\hat{S}}_i$, the nonzero elements of the permutation matrix are set to ${(1 - \sigma_i)}$, where $\sigma_i$ is the scattering coefficient of patch $i$.
The remaining elements are set to ${(\sigma_i / (M_i - 1))}$, where $M_i$ is the size of $\bm{\hat{S}}_i$.
With this configuration, the ratio of specular to diffuse energy is controlled by the scattering coefficient -- and the resulting matrix is doubly-stochastic, so the sign-agnostic optimization can be applied directly.

Note that this strategy allows for arbitrary scattering coefficients but not arbitrary \acp{BRDF}.
Also, note that neither this matrix design nor the Householder matrix design proposed in~\cite{arn} are equivalent to ideally diffuse scattering, since they redirect equal amounts of energy towards each patch regardless of the size, distance, and relative orientation of patches.

\subsection{High-order injection}
\label{sec:proposed-injection}

The second proposed extension is related to the injection operators, with the objective of enabling finer control of early reflections.
The injection operators are marked by $s_{ij}$ in \figurename~\ref{fig:fdn blocks}.
Note that we are here abusing notation, with the subscript in the figure indicating the delay line index, while the ones used in the text indicate the path between two patches.

As explained in Section~\ref{sec:background}, in~\cite{arn} the ``initial tracing'' step of \ac{ART} is used to evaluate the direct incidence radiance from the source, while in~\cite{fdn_art} it is used to evaluate the primary reflected radiance for each discrete path.
The former ensures accurate rendering only of the direct component, the latter only of the direct component and first-order reflections.

In our model, the injection operators are designed to inject radiance of order $K$ into the system.
That is to say, rather than basing their design on direct incidence radiance (see Eq.~\ref{eq:ART injection}), the radiance is reflected $K$ times before being assigned to delay lines.
Orders smaller or equal to $K$ are modeled via a filter (marked as $b$ in \figurename~\ref{fig:fdn blocks}) that bypasses the recursive part of the network.

A similar approach is commonly used in hybrid models involving the image model and \acp{FDN}~\cite{fifty_years}.
\acp{SDN} have been recently extended in a similar fashion~\cite{dafx_sdn, leny_sdn}, but are still limited to rectangular geometries.

\begin{figure}[t]
    \vspace*{-0.5cm}%
    \centering
    \subfloat[]{%
        \begin{adjustbox}{max width=0.24\textwidth}%
        \input{figures/injection_ord_1}%
        \end{adjustbox}%
        \label{fig:injection_first}%
    }
    \hfil
    \subfloat[]{%
        \begin{adjustbox}{max width=0.24\textwidth}%
        \input{figures/injection_ord_2}%
        \end{adjustbox}%
        \label{fig:injection_second}%
    }
    \caption{Evaluation of $\hat{L}_{0, ij} (n)$~(\ref{fig:injection_first}) and $\hat{L}_{1, ij} (n)$~(\ref{fig:injection_second}). The reflections shown here are strictly specular for visual clarity, while evaluation of network parameters may include scattering.}
    \label{fig:injection}
\end{figure}
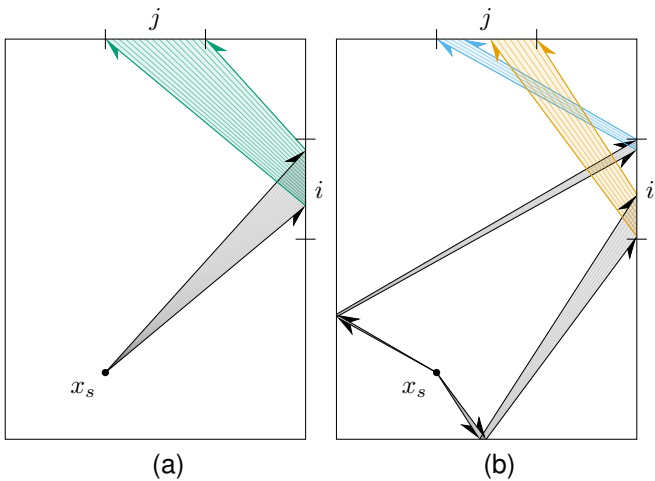

The injection radiance may be evaluated using ray-tracing, as is common practice for \ac{ART} and \ac{ARN}.
Unlike \ac{ART} and \ac{ARN}, however, the process we propose is carried out over reflection orders larger than one.
During the tracing process, each ray keeps memory of intersected patch indices, halting at the $(K+2)^\text{th}$ intersection.
Then, evaluation of order $K$ average radiance $\hat{L}_{K, ij} (n)$ is based on the rays which hit patch $i$ at their $(K+1)^\text{th}$ reflection and patch $j$ at the $(K+2)^\text{th}$.
The example in \figurename~\ref{fig:injection} shows the relevant ray bundles for delay line $ij$, with injection order ${K=0}$~(\ref{fig:injection_first}) or  ${K=1}$~(\ref{fig:injection_second}).

Two injector designs are considered here.
The first design is a simple combination of a delay line and a scaling coefficient, as used in the original \ac{ARN}~\cite{arn}.
The length of the delay line is calculated as the average travel time of the relevant rays, and the scaling coefficient is calculated as the square root of their average radiance.

A second approach is proposed here to enhance the modeled echo density, where the injectors operators are modeled using \ac{FIR} filters taking into account the temporal spreading of the ray bundle.
This is inspired by the approach used in~\cite{td-art} for recursive temporal spreading (note that we only add dispersion to the injection/detection operators, and not to recursive operators).
The gathered rays are first used to produce an energy impulse response~\cite[Chap.~11]{vorlander}.
The energy response is then divided by the total gathered radiance to obtain the required average, and finally its square root is used to modulate a noise signal, as is common practice in models based on squared pressure\cite{echogram_to_response}.
In the example of \figurename~\ref{fig:injection_second}, an injector including temporal spreading would correctly capture the two reflection paths as separate impulses, while a ``weighted delay'' injector would merge them into one.

Both of these methods enable the modeling of directional sources, by incorporating them into the ray tracing.

\subsection{Other parameters}
\label{sec:proposed-parameters}

In the following, we discuss the design of the remaining operators -- blocks $r_{ij}$, $d_{ij}$, $b$ in \figurename~\ref{fig:fdn blocks}.

\subsubsection{Detectors}

The detection operators $r_{ij}$ are evaluated similarly to the injection operators, by performing ray tracing, this time without performing any reflections.
The key difference is that the radiance of each bundle is not averaged.

Like the input operators, these can either be designed as simple scaled delays for simplicity, or include temporal spreading for richer echo density.
Receiver directional patterns are just as easily incorporated in the evaluation.
In particular, note that each detection operator $r_{ij}$ is related to a surface patch $j$, at a known distance and direction relative to the listener.
As such, the patches can be treated as individual audio objects for spatialization / binaural reproduction.
If so desired, the detection and bypass operators can include filters -- e.g. \acfp{HRTF} -- to carry out the spatial panning.
Simulation of microphone arrays is also possible, by modeling each pick-up point as a separate ``listener object'', and creating a separate set of detection operators for each one.

Since the output of injection operator $s_{ij}$ corresponds to the signal reaching patch $j$ from the direction of $i$, and the detection operator $r_{ij}$ relates that same quantity to the signal reaching the receiver from patch $j$, the two components in series produce a reflection of order ${K+1}$ ($s_{ij}$ performs ${K+1}$ reflections, $r_{ij}$ none).

\subsubsection{Delays}

The length of delay lines are based on the discrete delays $\bm{D}$.
In addition to the time delay, the $d_{ij}$ operators apply wall and air absorption, both of which may be frequency dependent.
Incorporating wall and air absorption in $d_{ij}$ is preferred over incorporating them in $\bm{A}$ to keep $\bm{A}$ unilossless.
This is easily achieved by use of normalized \acp{BRDF} and random-incidence absorption coefficients, as is common practice in \ac{GA}.

\subsubsection{Bypass}

The final component is the filter $b$, bypassing the recursive network.
It models reflections of orders ${\le K}$, including the line-of-sight component.
The simulations in the remainder of the paper employed \ac{ISM} to better model the impulsive nature of early reflections, but other methods can be used too.
For example, the operator may be evaluated during the same ray-tracing step as the injection operators, in which case it does not increase the computational complexity.

%% file: figures/injection_ord_1.tex
\begin{tikzpicture} [x=1, y=1, xscale=1.3, yscale=1.3]
    \definecolor{myOrange}{RGB}{230, 159, 0}; 
    \definecolor{myLightBlue}{RGB}{86, 180, 233}; 
    \definecolor{myGreen}{RGB}{0, 158, 115}; 
    \definecolor{myYellow}{RGB}{240, 228, 66}; 
    \definecolor{myDarkBlue}{RGB}{0, 114, 178}; 
    \definecolor{myRed}{RGB}{213, 94, 0}; 
    \definecolor{myPurple}{RGB}{204, 121, 167}; 
    
    \def \roomX {90};
    \def \roomY {120};
    \def \spacingJ {2};
    \def \opacity {0.1};
    
    \def \sourceX {30};
    \def \sourceY {20};
    
    \coordinate (Source) at (\sourceX, \sourceY);
    \coordinate (Image) at (2*\roomX - \sourceX, \sourceY);
    
    \coordinate (Corner_NW) at (0, \roomY);
    \coordinate (Corner_NE) at (\roomX, \roomY);
    \coordinate (Corner_SE) at (\roomX, 0);
    \coordinate (Corner_SW) at (0, 0);
    
    \coordinate (Patch_j_west) at (\roomX/3, \roomY);
    \coordinate (Patch_j_east) at (2*\roomX/3, \roomY);
    \coordinate (Patch_i_south) at (\roomX, 2*\roomY/4);
    \coordinate (Patch_i_north) at (\roomX, 3*\roomY/4);

    \coordinate (Patch_j_center) at ($(Patch_j_west)!0.5!(Patch_j_east)$);
    \coordinate (Patch_i_center) at ($(Patch_i_south)!0.5!(Patch_i_north)$);

    \path [overlay, name path=reflector_E] (Corner_SE) -- (Corner_NE);
    
    \edef \minray {-1};
    \edef \maxray {-1};
    \foreach \a in {0,\spacingJ,...,30}{
        \coordinate (j_point) at ([xshift=\a]Patch_j_west);
    
        \path [overlay, name path=image_path] (j_point) -- (Image);
        \path [overlay, name intersections={of=reflector_E and image_path, by=inters}];
        \newdimen\yOne;
        \newdimen\yTwo;
        \newdimen\yThree;
        \pgfextracty{\yOne}{\pgfpointanchor{inters}{center}};
        \pgfextracty{\yTwo}{\pgfpointanchor{Patch_i_south}{center}};
        \pgfextracty{\yThree}{\pgfpointanchor{Patch_i_north}{center}};
        \ifthenelse{\yTwo < \yOne}{
            \ifthenelse{\yOne < \yThree}{
                \ifthenelse{\minray < 0}{
                    \pgfmathparse {\a};
                    \xdef\minray {\pgfmathresult};
                }{};
                \ifthenelse{\maxray < \a}{
                    \pgfmathparse {\a};
                    \xdef\maxray {\pgfmathresult};
                }{};
                
                \draw [opacity=5*\opacity, draw=myGreen] (j_point) -- (inters);
                \draw [opacity=\opacity] (inters) -- (Source);
            }{};
        }{};
    };
    \coordinate (j_point_1) at ([xshift=\minray]Patch_j_west);
    \coordinate (j_point_2) at ([xshift=\maxray]Patch_j_west);
    \path [overlay, name path=image_path_1] (Image) -- (j_point_1);
    \path [overlay, name path=image_path_2] (Image) -- (j_point_2);
    \path [overlay, name intersections={of=reflector_E and image_path_1, by=inters_1}];
    \path [overlay, name intersections={of=reflector_E and image_path_2, by=inters_2}];
    
    \draw [fill=black, opacity=\opacity, draw=none] (Source) -- (inters_1) -- (inters_2) -- cycle;
    \draw [fill=myGreen, opacity=\opacity, draw=none] ([xshift=\maxray]Patch_j_west) -- ([xshift=\minray]Patch_j_west) -- (inters_1) -- (inters_2) -- cycle;

    \draw [arrows={-Stealth[harpoon, swap, length=3mm]}] (Source) -- (inters_1);
    \draw [arrows={-Stealth[harpoon, length=3mm]}] (Source) -- (inters_2);
    \draw [arrows={-Stealth[harpoon, swap, length=3mm]}, color=myGreen] (inters_2) -- ([xshift=\maxray]Patch_j_west);
    \draw [arrows={-Stealth[harpoon, length=3mm]}, color=myGreen] (inters_1) -- ([xshift=\minray]Patch_j_west);

    
    
    
    \draw (Corner_NW) -- (Corner_NE) -- (Corner_SE) -- (Corner_SW) -- cycle;
    
    \draw ([yshift=3]Patch_j_west) -- ([yshift=-3]Patch_j_west);
    \draw ([yshift=3]Patch_j_east) -- ([yshift=-3]Patch_j_east);
    \draw ([xshift=3]Patch_i_south) -- ([xshift=-3]Patch_i_south);
    \draw ([xshift=3]Patch_i_north) -- ([xshift=-3]Patch_i_north);
    
    \draw (Source) node [circle, fill, inner sep=1] {};
    \draw (Source) node [anchor=north east] {$x_s$};
    \draw (Patch_i_center) node [anchor=west] {$i$};
    \draw (Patch_j_center) node [anchor=south] {$j$};
\end{tikzpicture}

%% file: figures/injection_ord_2.tex
\begin{tikzpicture}[x=1, y=1, xscale=1.3, yscale=1.3]
    \definecolor{myOrange}{RGB}{230, 159, 0}; 
    \definecolor{myLightBlue}{RGB}{86, 180, 233}; 
    \definecolor{myGreen}{RGB}{0, 158, 115}; 
    \definecolor{myYellow}{RGB}{240, 228, 66}; 
    \definecolor{myDarkBlue}{RGB}{0, 114, 178}; 
    \definecolor{myRed}{RGB}{213, 94, 0}; 
    \definecolor{myPurple}{RGB}{204, 121, 167}; 
    
    \def \roomX {90};
    \def \roomY {120};
    \def \spacingJ {2};
    \def \opacity {0.1};
    
    \def \sourceX {30};
    \def \sourceY {20};
    
    \def \beamOne {16.3};
    \def \beamTwo {42.4};
    
    \coordinate (Source) at (\sourceX, \sourceY);
    \coordinate (Image_1_1) at (-\sourceX, \sourceY);
    \coordinate (Image_1_2) at (\sourceX + 2*\roomX, \sourceY);
    \coordinate (Image_2_1) at (\sourceX, -\sourceY);
    \coordinate (Image_2_2) at (2*\roomX - \sourceX, -\sourceY);
    
    \coordinate (Corner_NW) at (0, \roomY);
    \coordinate (Corner_NE) at (\roomX, \roomY);
    \coordinate (Corner_SE) at (\roomX, 0);
    \coordinate (Corner_SW) at (0, 0);
    
    \coordinate (Patch_j_west) at (\roomX/3, \roomY);
    \coordinate (Patch_j_east) at (2*\roomX/3, \roomY);
    \coordinate (Patch_i_south) at (\roomX, 2*\roomY/4);
    \coordinate (Patch_i_north) at (\roomX, 3*\roomY/4);

    \coordinate (Patch_j_center) at ($(Patch_j_west)!0.5!(Patch_j_east)$);
    \coordinate (Patch_i_center) at ($(Patch_i_south)!0.5!(Patch_i_north)$);

    \path [overlay, name path=reflector_E] (Corner_SE) -- (Corner_NE);
    \path [overlay, name path=reflector_W] (Corner_SW) -- (Corner_NW);
    \path [overlay, name path=reflector_S] (Corner_SW) -- (Corner_SE);
    
    \edef \minrayOne {-1};
    \edef \maxrayOne {-1};
    \foreach \a in {0,\spacingJ,...,30}{
        \coordinate (j_point) at ([xshift=\a]Patch_j_west);
    
        \path [overlay, name path=image_path_1_2] (j_point) -- (Image_1_2);
        \path [overlay, name intersections={of=reflector_E and image_path_1_2, by=inters_1_2}];
        \newdimen\yOne;
        \newdimen\yTwo;
        \newdimen\yThree;
        \pgfextracty{\yOne}{\pgfpointanchor{inters_1_2}{center}};
        \pgfextracty{\yTwo}{\pgfpointanchor{Patch_i_south}{center}};
        \pgfextracty{\yThree}{\pgfpointanchor{Patch_i_north}{center}};
        \ifthenelse{\yTwo < \yOne}{
            \ifthenelse{\yOne < \yThree}{
                \ifthenelse{\minrayOne < 0}{
                    \pgfmathparse {\a};
                    \xdef\minrayOne {\pgfmathresult};
                }{};
                \ifthenelse{\maxrayOne < \a}{
                    \pgfmathparse {\a};
                    \xdef\maxrayOne {\pgfmathresult};
                }{};
                \path [overlay, name path=image_path_1_1] (inters_1_2) -- (Image_1_1);
                \path [overlay, name intersections={of=reflector_W and image_path_1_1, by=inters_1_1}];
                
                \draw [opacity=5*\opacity, draw=myLightBlue] (j_point) -- (inters_1_2);
                \draw [opacity=\opacity] (inters_1_2) -- (inters_1_1);
                \draw [opacity=\opacity] (inters_1_1) -- (Source);
            }{};
        }{};
    };
    \coordinate (j_point_1) at ([xshift=\minrayOne]Patch_j_west);
    \coordinate (j_point_2) at ([xshift=\maxrayOne]Patch_j_west);
    \path [overlay, name path=image_path_1_2_1] (Image_1_2) -- (j_point_1);
    \path [overlay, name path=image_path_1_2_2] (Image_1_2) -- (j_point_2);
    \path [overlay, name intersections={of=reflector_E and image_path_1_2_1, by=inters_1_2_1}];
    \path [overlay, name intersections={of=reflector_E and image_path_1_2_2, by=inters_1_2_2}];
    \path [overlay, name path=image_path_1_1_1] (Image_1_1) -- (inters_1_2_1);
    \path [overlay, name path=image_path_1_1_2] (Image_1_1) -- (inters_1_2_2);
    \path [overlay, name intersections={of=reflector_W and image_path_1_1_1, by=inters_1_1_1}];
    \path [overlay, name intersections={of=reflector_W and image_path_1_1_2, by=inters_1_1_2}];
    
    \draw [fill=black, opacity=\opacity, draw=none] (Source) -- (inters_1_1_1) -- (inters_1_1_2) -- cycle;
    \draw [fill=black, opacity=\opacity, draw=none] (inters_1_1_1) -- (inters_1_1_2) -- (inters_1_2_2) -- (inters_1_2_1) -- cycle;
    \draw [fill=myLightBlue, opacity=\opacity, draw=none] ([xshift=\maxrayOne]Patch_j_west) -- ([xshift=\minrayOne]Patch_j_west) -- (inters_1_2_1) -- (inters_1_2_2) -- cycle;

    \draw [arrows={-Stealth[harpoon, length=3mm]}] (Source) -- (inters_1_1_1);
    \draw [arrows={-Stealth[harpoon, swap, length=3mm]}] (Source) -- (inters_1_1_2);
    \draw [arrows={-Stealth[harpoon, swap, length=3mm]}] (inters_1_1_1) -- (inters_1_2_1);
    \draw [arrows={-Stealth[harpoon, length=3mm]}] (inters_1_1_2) -- (inters_1_2_2);
    \draw [arrows={-Stealth[harpoon, swap, length=3mm]}, color=myLightBlue] (inters_1_2_2) -- ([xshift=\maxrayOne]Patch_j_west);
    \draw [arrows={-Stealth[harpoon, length=3mm]}, color=myLightBlue] (inters_1_2_1) -- ([xshift=\minrayOne]Patch_j_west);
    
    \edef \minrayTwo {-1};
    \edef \maxrayTwo {-1};
    \foreach \a in {0,\spacingJ,...,30}{
        \coordinate (j_point) at ([xshift=\a]Patch_j_west);
    
        \path [overlay, name path=image_path_2_2] (j_point) -- (Image_2_2);
        \path [overlay, name intersections={of=reflector_E and image_path_2_2, by=inters_2_2}];
        \newdimen\yOne;
        \newdimen\yTwo;
        \newdimen\yThree;
        \pgfextracty{\yOne}{\pgfpointanchor{inters_2_2}{center}};
        \pgfextracty{\yTwo}{\pgfpointanchor{Patch_i_south}{center}};
        \pgfextracty{\yThree}{\pgfpointanchor{Patch_i_north}{center}};
        \ifthenelse{\yTwo < \yOne}{
            \ifthenelse{\yOne < \yThree}{
                \ifthenelse{\minrayTwo < 0}{
                    \pgfmathparse {\a};
                    \xdef\minrayTwo {\pgfmathresult};
                }{};
                \ifthenelse{\maxrayTwo < \a}{
                    \pgfmathparse {\a};
                    \xdef\maxrayTwo {\pgfmathresult};
                }{};
                \path [overlay, name path=image_path_2_1] (inters_2_2) -- (Image_2_1);
                \path [overlay, name intersections={of=reflector_S and image_path_2_1, by=inters_2_1}];
                
                \draw [opacity=5*\opacity, draw=myOrange] (j_point) -- (inters_2_2);
                \draw [opacity=\opacity] (inters_2_2) -- (inters_2_1);
                \draw [opacity=\opacity] (inters_2_1) -- (Source);
            }{};
        }{};
    };
    \coordinate (j_point_1) at ([xshift=\minrayTwo]Patch_j_west);
    \coordinate (j_point_2) at ([xshift=\maxrayTwo]Patch_j_west);
    \path [overlay, name path=image_path_2_2_1] (Image_2_2) -- (j_point_1);
    \path [overlay, name path=image_path_2_2_2] (Image_2_2) -- (j_point_2);
    \path [overlay, name intersections={of=reflector_E and image_path_2_2_1, by=inters_2_2_1}];
    \path [overlay, name intersections={of=reflector_E and image_path_2_2_2, by=inters_2_2_2}];
    \path [overlay, name path=image_path_2_1_1] (Image_2_1) -- (inters_2_2_1);
    \path [overlay, name path=image_path_2_1_2] (Image_2_1) -- (inters_2_2_2);
    \path [overlay, name intersections={of=reflector_S and image_path_2_1_1, by=inters_2_1_1}];
    \path [overlay, name intersections={of=reflector_S and image_path_2_1_2, by=inters_2_1_2}];
    
    \draw [fill=black, opacity=\opacity, draw=none] (Source) -- (inters_2_1_1) -- (inters_2_1_2) -- cycle;
    \draw [fill=black, opacity=\opacity, draw=none] (inters_2_1_1) -- (inters_2_1_2) -- (inters_2_2_2) -- (inters_2_2_1) -- cycle;
    \draw [fill=myOrange, opacity=\opacity, draw=none] ([xshift=\maxrayTwo]Patch_j_west) -- ([xshift=\minrayTwo]Patch_j_west) -- (inters_2_2_1) -- (inters_2_2_2) -- cycle;

    \draw [arrows={-Stealth[harpoon, length=3mm]}] (Source) -- (inters_2_1_1);
    \draw [arrows={-Stealth[harpoon, swap, length=3mm]}] (Source) -- (inters_2_1_2);
    \draw [arrows={-Stealth[harpoon, swap, length=3mm]}] (inters_2_1_1) -- (inters_2_2_1);
    \draw [arrows={-Stealth[harpoon, length=3mm]}] (inters_2_1_2) -- (inters_2_2_2);
    \draw [arrows={-Stealth[harpoon, swap, length=3mm]}, color=myOrange] (inters_2_2_2) -- ([xshift=\maxrayTwo]Patch_j_west);
    \draw [arrows={-Stealth[harpoon, length=3mm]}, color=myOrange] (inters_2_2_1) -- ([xshift=\minrayTwo]Patch_j_west);

    \draw (Corner_NW) -- (Corner_NE) -- (Corner_SE) -- (Corner_SW) -- cycle;
    
    \draw ([yshift=3]Patch_j_west) -- ([yshift=-3]Patch_j_west);
    \draw ([yshift=3]Patch_j_east) -- ([yshift=-3]Patch_j_east);
    \draw ([xshift=3]Patch_i_south) -- ([xshift=-3]Patch_i_south);
    \draw ([xshift=3]Patch_i_north) -- ([xshift=-3]Patch_i_north);
    
    \draw (Source) node [circle, fill, inner sep=1] {};
    \draw (Source) node [anchor=north east] {$x_s$};
    \draw (Patch_i_center) node [anchor=west] {$i$};
    \draw (Patch_j_center) node [anchor=south] {$j$};
\end{tikzpicture}

%% file: sections/results.tex
\section{Evaluation}
\label{sec:results}

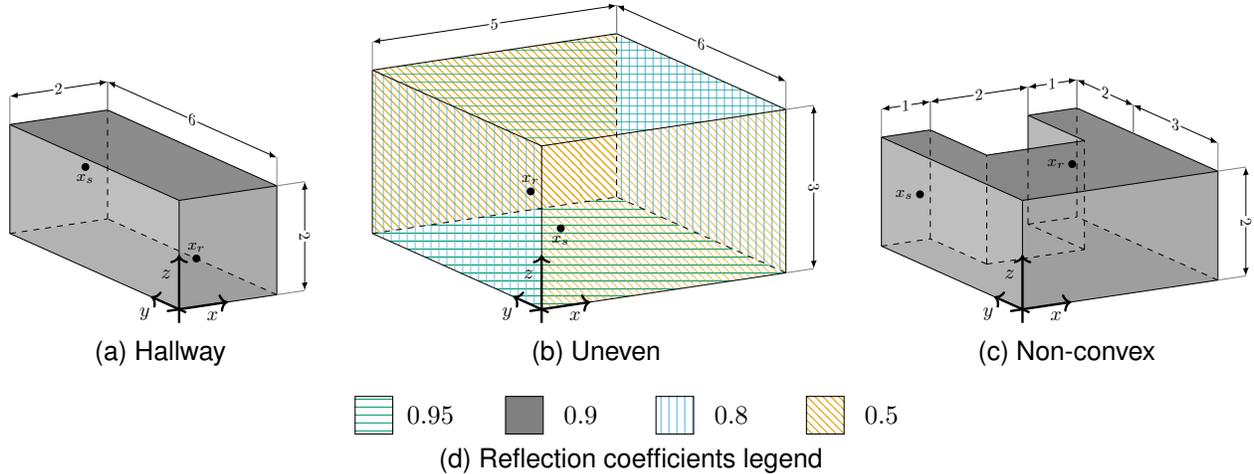
\begin{figure*}[!t]
    \input{figures/room_functions}
    \centering
    \subfloat[Hallway]{%
        \begin{adjustbox}{scale=0.75}%
        \input{figures/room_hallway}%
        \end{adjustbox}%
        \label{fig:rooms hallway}%
    }
    \hfil
    \subfloat[Uneven]{%
        \begin{adjustbox}{scale=0.75}%
        \input{figures/room_uneven}%
        \end{adjustbox}%
        \label{fig:rooms uneven}%
    }
    \hfil
    \subfloat[Non-convex]{%
        \begin{adjustbox}{scale=0.75}%
        \input{figures/room_non-convex}%
        \end{adjustbox}%
        \label{fig:rooms non-convex}%
    }
    \vfil
    \subfloat[Reflection coefficients legend]{%
        \begin{adjustbox}{max width=0.45\textwidth}%
        \begin{tikzpicture}
            \roomsLegend;
        \end{tikzpicture}%
        \end{adjustbox}%
        \label{fig:rooms legend}%
    }
    \caption{Geometry of the rooms presented in this paper, detailed in Table~\ref{tab:room parameters}.}
    \label{fig:rooms}
\end{figure*}

The proposed models were validated against two well-established \ac{GA} models (\acf{RTM} and \acf{ISM}), as well as real measurements.
The evaluation alongside \ac{GA} simulations was carried out to ensure a fair comparison, by considering simple scenarios with one defining characteristic (e.g.\@ unevenly distributed absorption or non-convex geometry, not both).
This granularity makes it possible to gauge each extension's contribution in relation to different properties of the scene.
The evaluation alongside real measurements acts as a more thorough validation of the method, in a more complex scenario with frequency-dependent and unevenly distributed absorption coefficients.
In both cases, comparisons were based on objective metrics, described in the following section.

Since our method is based on \ac{GA}, it inherits the limitations dictated by the acoustic ray approximation.
For this reason, it was deemed unnecessary to perform comparisons with wave-based simulations and/or examine the low-frequency performance in detail.

\subsection{Objective measures}

The reverberation time of each response was evaluated using octave-band $T_{30}$ values.
The performance in terms of the early response was quantified using \ac{EDT} and \ac{NED}.
These quantities are defined as follows.
\aclp{EDT} and $T_{30}$ are based on \acfp{EDC}, a measure of the energy remaining in a \ac{RIR} over time~\cite[Sec.~3.8]{kuttruff}, defined as
\begin{equation}
    \text{EDC}(n)
    =
    \frac{
        \sum_{i=n}^{\infty} h^{2}(i)
    }{
        \sum_{i=0}^{\infty} h^{2}(i)
    }
    \, ,
    \label{eq:edc}
\end{equation}
where $h(n)$ is a \ac{RIR}.
These can be evaluated for separate frequency bands as well as for the full, broadband \ac{RIR}.
The $T_{60}$ value, measured in seconds, is defined as the moment the \ac{EDC} passes \qty{-60}{\decibel}; $T_{30}$ shares that definition, but is computed based on a linear regression of the \ac{EDC} in the range \qtyrange[range-units = repeat]{-5}{-35}{\decibel}.
\acl{EDT} is defined as the moment the \ac{EDC} passes \qty{-10}{\decibel}.

The \acf{NED} is a measure devised to estimate the perceived sense of sound diffusion in a space~\cite{ned}.
It is defined, for a given time window of the \ac{RIR}, as the ratio between the number of samples lying more than a standard deviation away from the mean and that which is expected for Gaussian noise.

\subsection{Simulation comparisons}

The \ac{GA} models used as references are CATT-Acoustics~\cite{catt_acoustics} (a state-of-the-art \ac{RTM} model) and PyRoomAcoustics~\cite{pyroomacoustics} (an \ac{ISM} model).
The \ac{ISM} does not model scattering and is only shown as a reference for a single scenario with low scattering.

Simulation examples included three feedback matrix designs (Householder, Sinkhorn-Knopp, uniform), both injection operator designs (with or without time spreading), several injection orders $K$, and several surface discretization levels.
Results marked as ``\acs{ARN} baseline'' use Householder matrix designs and injection operators with ${K=1}$ and no time spreading, combining the advantages of \ac{ARN}~\cite{arn} and \acs{FDN}-\acs{ART}~\cite{fdn_art}.

Reflection kernels were evaluated based on the common ``pseudospecular'' \ac{BRDF}~\cite{rare}, performing the integral in Eq.~(\ref{eq:discretized kernel}) by sampling each reflector at \qty{0.5}{\meter} intervals and performing ray-tracing Monte Carlo integration from each sample point.
Injection and detection operators were evaluated using ray-tracing as previously described, while the bypass filter was evaluated separately using \ac{ISM}.

Simulated environments included two rectangular rooms and one non-convex room, all of which were tested for different source-receiver configurations, scattering coefficients, and wall absorption coefficients, including scenarios with unevenly-distributed absorption.
All simulations used the same air absorption parameters (\qty{50}{\percent} humidity, \qty{20}{\degreeCelsius}) and omnidirectional source and receiver.
All responses were processed with a \qty{20}{\hertz} high-pass filter.
Note that all reported octave-band plots start from \qty{125}{\hertz} center frequency, as that is the lowest frequency band modeled by CATT-Acoustics.

\begin{table}[tb]
    \caption{\itshape Parameters used for the room simulations presented in this paper. For the ``non-convex'' room, the size reported here is that of the bounding box.}
    \centering
    \renewcommand{\arraystretch}{1.2}
    \begin{tabular}{|cc||c|c|c|}
        \hline
        \multicolumn{2}{|c||}{Descriptor} & Hallway & Non-convex & Uneven\\
        \hline
         & $x$ & \qty{2}{\metre} & \qty{4}{\metre} & \qty{5}{\metre} \\
        Size & $y$ & \qty{6}{\metre} & \qty{5}{\metre} & \qty{6}{\metre} \\
         & $z$ & \qty{2}{\metre} & \qty{2}{\metre} & \qty{3}{\metre} \\
        \hline
         & $x$ & \qty{1.2}{\metre} & \qty{0.5}{\metre} & \qty{1.2}{\metre} \\
        Source pos. & $y$ & \qty{5.4}{\metre} & \qty{4.5}{\metre} & \qty{1.4}{\metre} \\
         & $z$ & \qty{1.2}{\metre} & \qty{1.0}{\metre} & \qty{1.0}{\metre} \\
        \hline
         & $x$ & \qty{0.7}{\metre} & \qty{3.5}{\metre} & \qty{0.7}{\metre} \\
        Mic pos. & $y$ & \qty{0.6}{\metre} & \qty{4.3}{\metre} & \qty{1.6}{\metre} \\
         & $z$ & \qty{0.7}{\metre} & \qty{1.2}{\metre} & \qty{1.7}{\metre} \\
        \hline
        \multicolumn{2}{|c||}{Reflection coeff.} & 0.9 & 0.9 & 0.95, 0.8, 0.5 \\
        \hline
        \multicolumn{2}{|c||}{Scattering coeff.} & 0.25 & 0.25 & 0.05 \\
        \hline
    \end{tabular}
    \label{tab:room parameters}
\end{table}

Only some salient examples are reported here, while all setups and responses are available online~\cite{rir_repository}.
The environments presented in the following are detailed in Table~\ref{tab:room parameters}, and illustrated in \figurename~\ref{fig:rooms}.

Note that the employed wall absorption coefficients are frequency-independent, and the high-frequency roll-off observed in the results is entirely due to air absorption.
Frequency-dependent wall absorption coefficients can easily be employed as part of the delay operators as discussed in the previous section, but were not included here for simplicity.

\subsubsection{Reverberation time}

\figurename~\ref{fig:t30 1} shows the octave-band $T_{30}$ values in the room with uneven absorption.
The first observation to be made here relates to the baseline ARN model, which shows an over-estimation of the $T_{30}$ at all frequencies.
Crucially, the overestimation worsens with increased spatial discretization.
Matrix sizes of ${M=208}$ correspond to a discretization with ${N=16}$ patches, ${M=1458}$ to ${N=42}$ patches, and ${M=3328}$ to ${N=64}$ patches.
This is due to the fact that in the baseline ARN, the scattering matrix is given by a Householder transformation whereby the specular-to-diffuse energy ratio depends on the matrix size. 
This is clearly problematic, since it gives rise to a trade-off between spatial accuracy and $T_{30}$ accuracy.

\begin{figure}[tb]
    \begin{center}
    \begin{adjustbox}{max width=0.48\textwidth}
    \input{figures/T30_5}
    \end{adjustbox}
    \end{center}
    \caption{Octave-band $T_{30}$ values for the ``uneven'' room. All \acp{ARN} use basic injectors with ${K=1}$ and no time dispersion.
    }
    \label{fig:t30 1}
\end{figure}
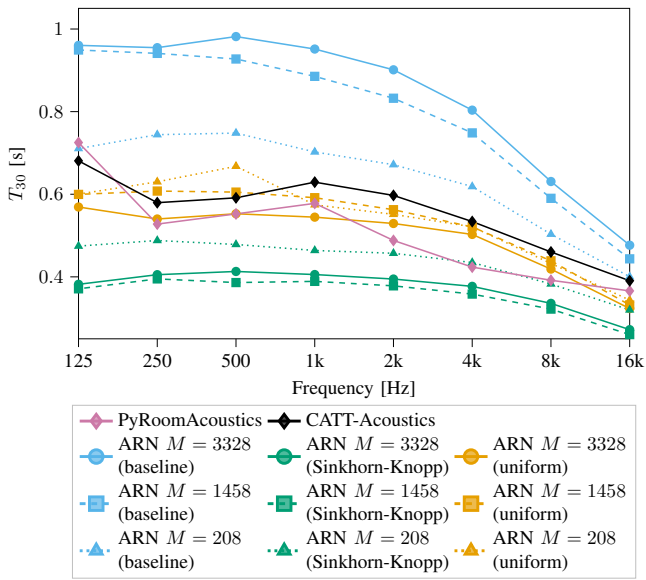

\begin{figure}[tb]
    \begin{center}
    \begin{adjustbox}{max width=0.48\textwidth}
    \input{figures/T30_4}
    \end{adjustbox}
    \end{center}
    \caption{Octave-band $T_{30}$ values for the ``non-convex'' room. All \acp{ARN} use basic injectors with ${K=1}$ and no time dispersion.
    }
    \label{fig:t30 2}
\end{figure}
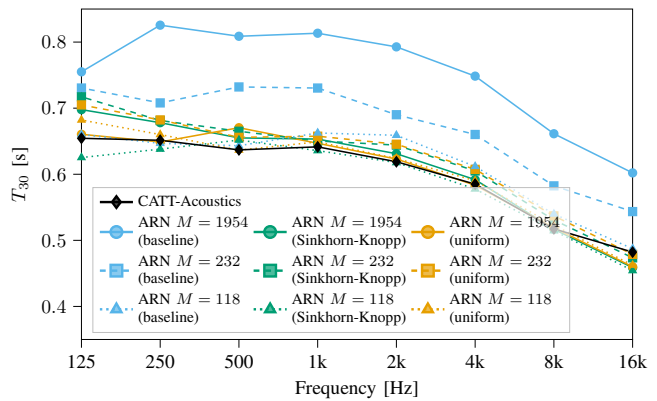

The proposed ``uniform'' design, on the other hand, produces reliably accurate reverberation times, with minimal variation when the matrix sizes change.
The proposed ``Sinkhorn-Knopp'' design results in an underestimation of $T_{30}$ values, but is still closer to CATT-Acoustics compared to the baseline.

The large differences in $T_{30}$ obtained with different feedback matrix designs suggest that, in environments with uneven absorption such as the one shown, scattering plays an important role in reverberation times.
This is supported by results not presented here for space reasons, which show that CATT-Acoustics produces different $T_{30}$ values for different scattering coefficients.
Said difference is most pronounced in the uneven room, but also appears in the non-convex one.

Householder matrices can also lead to incorrect $T_{30}$ values in some uniform absorption cases, such as the non-convex room.
\figurename~\ref{fig:t30 2} shows the $T_{30}$ results for said room.
Like for the uneven-absortion room, CATT-Acoustics suggest an influence of the scattering coefficients on reverberation times for the non-convex room, although it was less pronounced in this case.
In this instance, both of the proposed designs perform well, matching the reference (CATT-Acoustics) very closely regardless of the discretization level.
In this case, ${M=118}$ corresponds to a discretization with ${N=14}$ patches, ${M=232}$ to ${N=19}$ patches, and ${M=1954}$ to ${N=58}$ patches.

In rectangular rooms with uniform absorption coefficients, such as the ``hallway'' room, all \acp{ARN} -- including the baseline -- generally produce correct $T_{30}$ values.
Such environments have been tested in previous works employing Householder matrices~\cite{arn, efficient_sdn, dafx_sdn, leny_sdn}.

\subsubsection{Early decay time}

\begin{figure}[tb]
    \begin{center}
    \begin{adjustbox}{max width=0.48\textwidth}
    \input{figures/EDT_2}
    \end{adjustbox}
    \end{center}
    \caption{Octave-band \ac{EDT} values for the ``non-convex'' room. The ``baseline'' \ac{ARN} uses a Householder matrix as usual, while the proposed ones use the uniform matrix design. The matrix size for all \acp{ARN} is ${M=1954}$, corresponding to ${N=58}$ patches.
    }
    \label{fig:edt 2}
\end{figure}
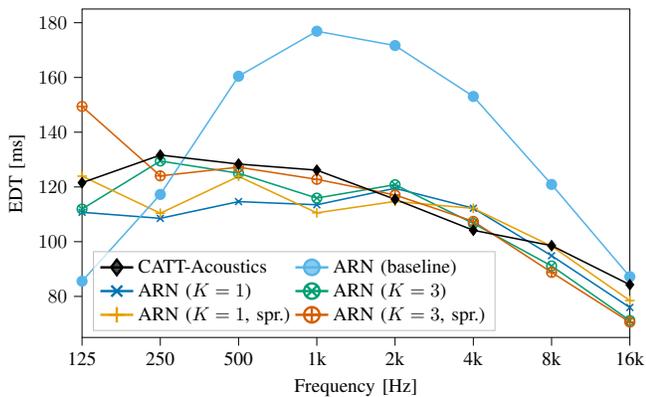

Householder matrices with a large size can also cause inaccurate modeling of \acf{EDT}.
More specifically, they can result in an abrupt drop in energy between early and late reflections, an effect previously observed in~\cite{efficient_sdn, dafx_sdn, wgw, hacihabiboglu2011frequency}, regardless of the room shape and the absorption coefficients.
Both of the proposed matrix designs prevent this effect.
\figurename~\ref{fig:edt 2} shows examples of octave-band \ac{EDT} values for different matrix designs and injector designs.
Householder matrices (shown only for the ``baseline'') lead to excessive \ac{EDT} for all frequency bands above \qty{500}{Hz}.
The ``uniform'' matrix design leads to more accurate values regardless of the employed injector design; ``Sinkhorn-Knopp'' design results (not shown in \figurename~\ref{fig:edt 2}) were similarly accurate.
With either of the proposed matrix designs, a slight improvement in accuracy can be seen when the injection order is raised from ${K=1}$ to ${K=3}$.
Note that introducing temporal spreading in the injector design has minimal effects on the \ac{EDT}.

\subsubsection{Early reflections}

\begin{figure}[tb]
    \begin{center}
    \begin{adjustbox}{max width=0.48\textwidth}
    \input{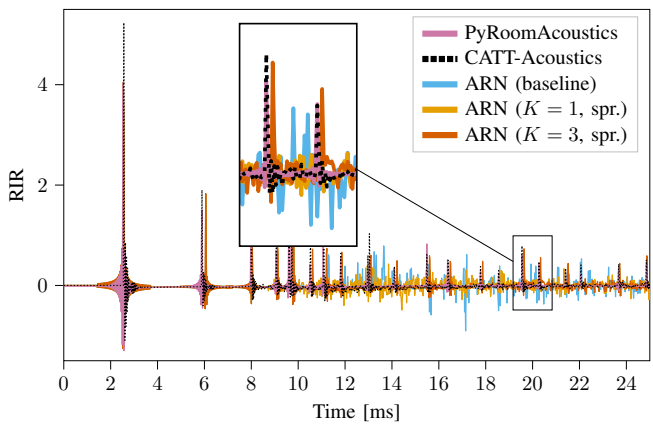}
    \end{adjustbox}
    \end{center}
    \caption{\acl{RIR} for the ``uneven'' room comparing the injection operators with and without temporal spreading, and for different injection orders, $K$. The ``baseline'' \ac{ARN} uses a Householder matrix as usual, while the proposed ones use the uniform matrix design. The matrix size for all \acp{ARN} is ${M=3328}$, corresponding to ${N=64}$ patches.
    }
    \label{fig:rir}
\end{figure}

\begin{figure}[tb]
    \begin{center}
    \begin{adjustbox}{max width=0.48\textwidth}
    \input{figures/EDC_1}
    \end{adjustbox}
    \end{center}
    \caption{Early section of the \acl{EDC} for the ``hallway'' room. Comparison of injectors with or without temporal spreading, $K$ marks the injection order. All \acp{ARN} use a Householder matrix. The matrix size for all \acp{ARN} is ${M=30}$, corresponding to ${N=6}$ patches.
    }
    \label{fig:edc}
\end{figure}

Increasing the injection order $K$ improves the accuracy of prominent early reflections.
This can be seen in the \acp{RIR} in \figurename~\ref{fig:rir}.
The inset plot highlights a part of the \ac{RIR} where the proposed method with ${K=3}$ produces exact reflections while the baseline and proposed method with ${K=1}$ do not.

\figurename~\ref{fig:edc} further illustrates the increase in accuracy of the early response, in the form of \acp{EDC}.
The illustrated ``hallway'' room is a rectangular room with uniform absorption coefficients, a scenario where the baseline produces correct reverberation times.
This was chosen to ensure a fair comparison, since the different decay in other rooms affects the early portion of the \ac{EDC} as well.
Moreover, the presented case is one with very coarse surface discretization, for reasons explained in the next subsection.

\figurename~\ref{fig:edc} shows how increasing the injection order $K$ improves the profile of the early decay, resulting in improved \ac{EDT} prediction (as in \figurename~\ref{fig:edt 2}).
It can be observed that, as anticipated by the \acp{EDT}, the presence of temporal spreading has negligible effect on the late energy decay.

\figurename~\ref{fig:edc} also shows certain artefacts resulting from the coarse discretization of the boundary.
For instance, in the ${K=1}$ case, the \ac{RIR} onset begins even earlier than the line-of-sight component.
This is due to the fact that early signals traveling through the delay lines can find much shorter paths than would be expected.
This is particularly evident in cases like the one shown here, where the source and receiver are positioned at opposite ends of the elongated space and lengthy surfaces are modeled by individual patches.
Increasing $K$ does not remove these excessively short paths from the system, but it delays the onset of this type of artefacts.

It should be noted that increasing the spatial discretization reduces these artefacts, but at a significant computational cost.
The proposed higher-order injection method, on the other hand, allows accurate control of the early reflections with only a small impact on the computational complexity.

\begin{figure}[tb]
    \begin{center}
    \begin{adjustbox}{max width=0.48\textwidth}
    \input{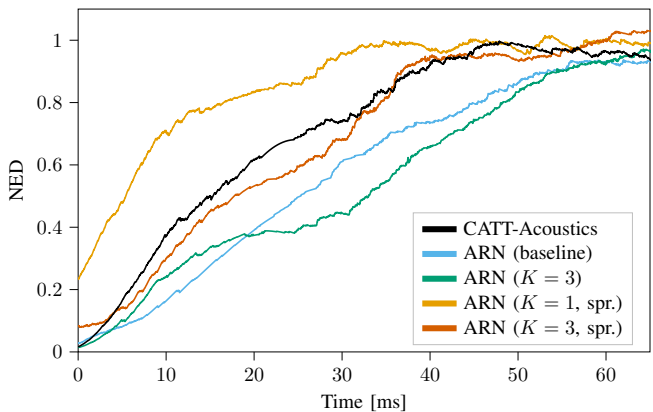}
    \end{adjustbox}
    \end{center}
    \caption{\acs{NED} for the ``hallway'' room. Comparison of injectors with or without temporal spreading, $K$ marks the injection order. \acs{NED} window length \qty{25}{\milli\second}. All \acp{ARN} use a Householder matrix. The matrix size for all \acp{ARN} is ${M=30}$, corresponding to ${N=6}$ patches.
    }
    \label{fig:ned}
\end{figure}

\subsubsection{Normalized Echo density}

\figurename~\ref{fig:ned} shows the time evolution of \ac{NED} produced with the same simulation setup as \figurename~\ref{fig:edc}.
Two effects can be observed for the case ${K=3}$ with no temporal spreading.
First, the very early echo density is closer than the baseline to CATT-Acoustics, thanks to the bypass filter producing correct early reflections.
Then, after about \si{15\milli\second}, the density decreases below the baseline: this is due the fact that with large surface patches, groups of high-order reflections get ``bundled'' together (e.g.\@ see \figurename~\ref{fig:injection_second}).
Using ${K=3}$ injectors with temporal spreading resolves this issue, because multiple reflections in one injector lead to an increase in echo density, getting closer to CATT-Acoustics.

It should be noted, however, that temporal spreading can cause an excessive echo density when used with injectors of low order.
For instance, \figurename~\ref{fig:ned} shows the ${K=1}$ case with temporal spreading, which results in a very high echo density being produced too soon.
This is due the fact that with very large surface patches, the recursive delay lines lead to premature diffusion, as discussed in the previous sub-section.
When combined with the ``noisy'' temporal spreading design, this makes the echo density rise excessively fast.
A higher injection order, like ${K=3}$, prevents this by design.

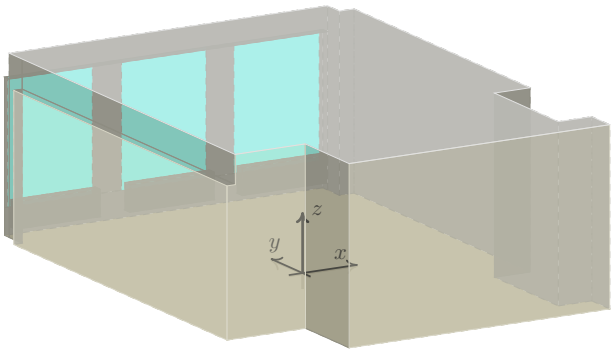
\begin{figure}[tb]
    \centering
    \begin{adjustbox}{max width=0.45\textwidth}
    \input{figures/cr2_really_simplified}
    \end{adjustbox}
    \caption{The simplified model of scene CR2~\cite{BRAS} used for validation. Note that some details of the original model were removed (ceiling lights, small protrusions on right wall).}
    \label{fig:BRAS room}
\end{figure}

\subsection{Measurement comparisons}

The CR2 scene from the \acf{BRAS} dataset~\cite{BRAS} was employed for the comparison against ground-truth measurements.
The model's geometry was slightly simplified (as shown in \figurename~\ref{fig:BRAS room}) in order to reduce the number of polygons with small area.
The parameters provided in the dataset were used to model frequency-dependent air and wall\footnote{The ``fitted'' absorption coefficients were employed, see~\cite{BRAS} for details.} absorption, however the scattering was modeled with frequency-independent coefficients, since the unilossless optimization of polynomial matrices is still an open problem.
Each material was modeled using the provided scattering coefficient at \qty{4}{\kilo\hertz}.

The measurements made using the dodecahedral speaker were chosen as reference, and the speaker was modeled as a single point source at the position of the high-frequency unit.
All loudspeaker and microphone positions were tested.
The results presented here are related to the pair LS1-MP2, but the reported observations apply to all configurations.

The loudspeakers' and microphones' directional transfer functions were not applied.
As a result, the simulated responses' normalization might be lacking a scaling factor, but this is inconsequential to all considered metrics (i.e. $T_{30}$, \ac{EDT}, \ac{NED}).

As with the \ac{GA} comparisons, the method was tested with different matrix and injector designs, injection orders, and surface discretization levels.
Changing the level of surface discretization had similar effects to those discussed above. 
For brevity, results are only reported for one discretization setup, featuring ${N=82}$ patches and ${M=4108}$ delay lines.

\subsubsection{Reverberation time}

\begin{figure}[tb]
    \begin{center}
    \begin{adjustbox}{max width=0.48\textwidth}
    \input{figures/T30_BRAS}
    \end{adjustbox}
    \end{center}
    \caption{Octave-band $T_{30}$ comparison against the real measurement in scene CR2, configuration LS1-MP2~\cite{BRAS}. All \acp{ARN} except the baseline use a Sinkhorn-Knopp matrix design. The matrix size for all \acp{ARN} is ${M=4108}$, corresponding to ${N=82}$ patches.
    }
    \label{fig:BRAS t30}
\end{figure}
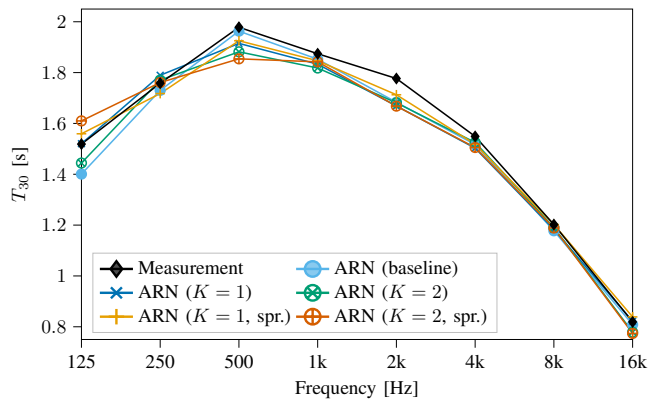

All reverberation times are in very good accordance with the ground truth, as shown in \figurename~\ref{fig:BRAS t30}.
Employing different matrix designs seems to have negligible effects on the $T_{30}$.
This is likely due to the room being mostly convex, and featuring uneven but relatively uniform absorption coefficients.
The injector order and design also has minimal effect on the reverberation times, which is expected.

\subsubsection{Early reverberation}

\begin{figure}[tb]
    \begin{center}
    \begin{adjustbox}{max width=0.48\textwidth}
    \input{figures/EDT_BRAS}
    \end{adjustbox}
    \end{center}
    \caption{Octave-band \ac{EDT} comparison against the real measurement in scene CR2, configuration LS1-MP2~\cite{BRAS}. All \acp{ARN} except the baseline use a Sinkhorn-Knopp matrix design. The matrix size for all \acp{ARN} is ${M=4108}$, corresponding to ${N=82}$ patches.
    }
    \label{fig:BRAS edt}
\end{figure}
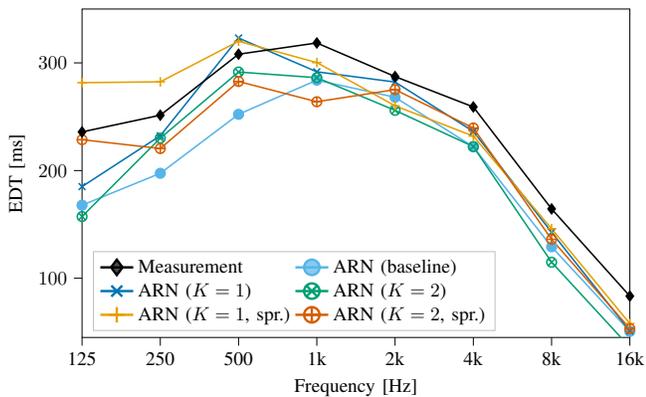

\begin{figure}[tb]
    \begin{center}
    \begin{adjustbox}{max width=0.48\textwidth}
    \input{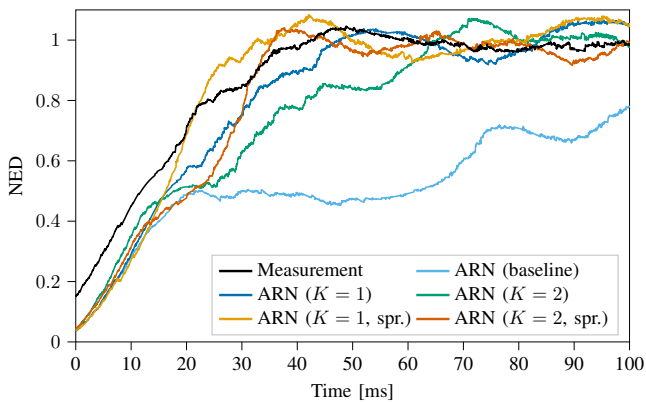}
    \end{adjustbox}
    \end{center}
    \caption{\acs{NED} comparison against the real measurement in scene CR2, configuration LS1-MP2~\cite{BRAS}. \acs{NED} window length \qty{25}{\milli\second}. All \acp{ARN} except the baseline use a Sinkhorn-Knopp matrix design. The matrix size for all \acp{ARN} is ${M=4108}$, corresponding to ${N=82}$ patches.
    }
    \label{fig:BRAS ned}
\end{figure}

The measured responses exhibit relatively early mixing times, most likely due to the intricate details of several walls in the room.
On one hand, this means that increasing the injection order and design has little effect on the \ac{EDT}, as shown in \figurename~\ref{fig:BRAS edt}.
On the other hand, the injector design is quite relevant in terms of echo density (see \figurename~\ref{fig:BRAS ned}), since the application of temporal spreading accelerates the mixing.

All \acp{ARN} exhibited a small underestimation of the \ac{EDT}, as can be seen in \figurename~\ref{fig:BRAS edt}.
This is to be expected, as was discussed in the previous section (refer back to \figurename~\ref{fig:edc}).
The same behaviour was observed for all matrix and injector designs.

\section{Computational Complexity}
\label{sec:results-complexity}

The computational complexity of this class of models can be separated between a precomputation cost, an interactive update cost, and an audio processing cost.
This section explains what the different components amount to, and provides a qualitative description of the additional cost associated to the proposed extensions compared to the baseline \acp{ARN}~\cite{arn}.

\subsection{Precomputation cost}

The precomputation cost includes the calculation of the scattering matrix $\bm{A}$ and of the recursion operators $d_{ij}$.
This is the component that is most computationally heavy, as it involves ray-tracing from a high number of surface points.
Unless the surface geometry or material properties change over time, these parameters need to be computed only once.

Compared to the baseline \acp{ARN}, the matrix designs proposed in this paper also require finding the ``closest unilossless'' matrix as described earlier.
However, the cost of this operation is negligible compared to that of evaluating $\bm{\hat{S}}$, which is always required.
The recursion operators $d_{ij}$ are unchanged with respect to the baseline.

\subsection{Interactive update cost}

The injection, detection, and bypass operators ($s_{ij}$, $r_{ij}$, and $b$, respectively), depend on the positions of sources and listeners, and therefore need to be recomputed any time they move.
More specifically, the injection operators require updating for source movements, the detection operators for receiver movements, and the bypass filter for either.
Their computation, as detailed in the previous section, involves ${K+2}$ ray-tracing reflections for injectors, and one for detectors.

Compared to the baseline \acp{ARN}, the design cost of the injection operators $s_{ij}$ rises linearly with the desired injection order $K$, due to the additional ray-tracing steps.
The detection operators $r_{ij}$ are essentially unchanged with respect to the baseline.
If the \ac{ISM} is employed for the bypass operator $b$, the associated computational cost rises quadratically (or, in the case of non-rectangular rooms, exponentially) with the desired injection order $K$.
If the \ac{RTM} is employed instead, there is no additional cost associated to $b$, since it is evaluated at the same time as $s_{ij}$ (as discussed in the previous section).
Note that a hybrid approach can be taken: for example, if the desired injection order is ${K=3}$, one could employ \ac{ISM} for first order reflections, and \ac{RTM} for second and third order ones.

\subsection{Audio processing cost}

The audio processing cost involves running the filter structure as shown in \figurename~\ref{fig:fdn blocks}, once all parameters are set.
Compared to the baseline \acp{ARN}, the additional cost of the proposed extensions comes from the filter operation of the bypass operator $b$ and the injection/detection operators $s_{ij}$ and $r_{ij}$ (if using the temporal spreading method).
The feedback matrix $\bm{A}$ has the form of a block-matrix for both the baseline \acp{ARN} and the proposed extension.
However, as opposed to the baseline \acp{ARN}, sub-matrices in the proposed extension are no longer Householder matrices.
This means that the computational cost for each block increases from $O(M_i)$ to $O(M_i^2)$, where $M_i$ is the size of the block associated to the~$i^\text{th}$ patch.

%% file: figures/room_functions.tex
\newcommand\shoebox[9]{%
    \def \X {#1};
    \def \Y {#2};
    \def \Z {#3};
    \def \Opacity {0.8};
    
    \fill[#4] (tpp cs:x=0,y=0,z=0)
    -- (tpp cs:x=0,y=\Y,z=0)
    -- (tpp cs:x=\X,y=\Y,z=0)
    -- (tpp cs:x=\X,y=0,z=0) -- cycle;
    \fill[#6]  (tpp cs:x=\X,y=0,z=0)
    -- (tpp cs:x=\X,y=0,z=\Z)
    -- (tpp cs:x=\X,y=\Y,z=\Z)
    -- (tpp cs:x=\X,y=\Y,z=0) -- cycle;
    \fill[#8] (tpp cs:x=0,y=\Y,z=0)
    -- (tpp cs:x=0,y=\Y,z=\Z)
    -- (tpp cs:x=\X,y=\Y,z=\Z)
    -- (tpp cs:x=\X,y=\Y,z=0) -- cycle;
    \fill[#5, opacity=\Opacity] (tpp cs:x=0,y=0,z=\Z)
    -- (tpp cs:x=0,y=\Y,z=\Z)
    -- (tpp cs:x=\X,y=\Y,z=\Z)
    -- (tpp cs:x=\X,y=0,z=\Z) -- cycle;
    \fill[#7, opacity=\Opacity]  (tpp cs:x=0,y=0,z=0)
    -- (tpp cs:x=0,y=0,z=\Z)
    -- (tpp cs:x=0,y=\Y,z=\Z)
    -- (tpp cs:x=0,y=\Y,z=0) -- cycle;
    \fill[#9, opacity=\Opacity] (tpp cs:x=0,y=0,z=0)
    -- (tpp cs:x=0,y=0,z=\Z)
    -- (tpp cs:x=\X,y=0,z=\Z)
    -- (tpp cs:x=\X,y=0,z=0) -- cycle;

	\draw[thin] (0,0,0) -- (\X,0,0);
	\draw[thin] (0,0,\Z) -- (\X,0,\Z);
	\draw[thin, dashed] (0,\Y,0) -- (\X,\Y,0);
	\draw[thin] (0,\Y,\Z) -- (\X,\Y,\Z);
	\draw[thin] (0,0,0) -- (0,\Y,0);
	\draw[thin, dashed] (\X,0,0) -- (\X,\Y,0);
	\draw[thin] (0,0,\Z) -- (0,\Y,\Z);
	\draw[thin] (\X,0,\Z) -- (\X,\Y,\Z);
	\draw[thin] (0,0,0) -- (0,0,\Z);
	\draw[thin] (\X,0,0) -- (\X,0,\Z);
	\draw[thin] (0,\Y,0) -- (0,\Y,\Z);
	\draw[thin, dashed] (\X,\Y,0) -- (\X,\Y,\Z);
}

\newcommand\nonconvexC[5]{%
    \def \Xone {#1};
    \def \Xtwo {\the\numexpr #1 + #2 \relax};
    \def \Xthree {\the\numexpr #1 + #2 + #1 \relax};
    \def \Yone {#3};
    \def \Ytwo {\the\numexpr #3 + #4 \relax};
    \def \Z {#5};
    \def \Opacity {0.8};
    
    \fill[gray] (tpp cs:x=0,y=0,z=0)
    -- (tpp cs:x=0,y=\Ytwo,z=0)
    -- (tpp cs:x=\Xone,y=\Ytwo,z=0)
    -- (tpp cs:x=\Xone,y=\Yone,z=0)
    -- (tpp cs:x=\Xtwo,y=\Yone,z=0)
    -- (tpp cs:x=\Xtwo,y=\Ytwo,z=0)
    -- (tpp cs:x=\Xthree,y=\Ytwo,z=0)
    -- (tpp cs:x=\Xthree,y=0,z=0) -- cycle;
    \fill[gray!50!white]  (tpp cs:x=\Xthree,y=0,z=0)
    -- (tpp cs:x=\Xthree,y=0,z=\Z)
    -- (tpp cs:x=\Xthree,y=\Ytwo,z=\Z)
    -- (tpp cs:x=\Xthree,y=\Ytwo,z=0) -- cycle;
    \fill[gray!50!white]  (tpp cs:x=\Xone,y=\Yone,z=0)
    -- (tpp cs:x=\Xone,y=\Yone,z=\Z)
    -- (tpp cs:x=\Xone,y=\Ytwo,z=\Z)
    -- (tpp cs:x=\Xone,y=\Ytwo,z=0) -- cycle;
    \fill[gray!70!white] (tpp cs:x=0,y=\Ytwo,z=0)
    -- (tpp cs:x=0,y=\Ytwo,z=\Z)
    -- (tpp cs:x=\Xone,y=\Ytwo,z=\Z)
    -- (tpp cs:x=\Xone,y=\Ytwo,z=0) -- cycle;
    \fill[gray!70!white] (tpp cs:x=\Xtwo,y=\Ytwo,z=0)
    -- (tpp cs:x=\Xtwo,y=\Ytwo,z=\Z)
    -- (tpp cs:x=\Xthree,y=\Ytwo,z=\Z)
    -- (tpp cs:x=\Xthree,y=\Ytwo,z=0) -- cycle;
    \fill[gray!50!white, opacity=\Opacity]  (tpp cs:x=\Xtwo,y=\Yone,z=0)
    -- (tpp cs:x=\Xtwo,y=\Yone,z=\Z)
    -- (tpp cs:x=\Xtwo,y=\Ytwo,z=\Z)
    -- (tpp cs:x=\Xtwo,y=\Ytwo,z=0) -- cycle;
    \fill[gray!70!white] (tpp cs:x=\Xone,y=\Yone,z=0)
    -- (tpp cs:x=\Xone,y=\Yone,z=\Z)
    -- (tpp cs:x=\Xtwo,y=\Yone,z=\Z)
    -- (tpp cs:x=\Xtwo,y=\Yone,z=0) -- cycle;
    \fill[gray, opacity=\Opacity] (tpp cs:x=0,y=0,z=\Z)
    -- (tpp cs:x=0,y=\Ytwo,z=\Z)
    -- (tpp cs:x=\Xone,y=\Ytwo,z=\Z)
    -- (tpp cs:x=\Xone,y=\Yone,z=\Z)
    -- (tpp cs:x=\Xtwo,y=\Yone,z=\Z)
    -- (tpp cs:x=\Xtwo,y=\Ytwo,z=\Z)
    -- (tpp cs:x=\Xthree,y=\Ytwo,z=\Z)
    -- (tpp cs:x=\Xthree,y=0,z=\Z) -- cycle;
    \fill[gray!50!white, opacity=\Opacity]  (tpp cs:x=0,y=0,z=0)
    -- (tpp cs:x=0,y=0,z=\Z)
    -- (tpp cs:x=0,y=\Ytwo,z=\Z)
    -- (tpp cs:x=0,y=\Ytwo,z=0) -- cycle;
    \fill[gray!70!white, opacity=\Opacity] (tpp cs:x=0,y=0,z=0)
    -- (tpp cs:x=0,y=0,z=\Z)
    -- (tpp cs:x=\Xthree,y=0,z=\Z)
    -- (tpp cs:x=\Xthree,y=0,z=0) -- cycle;

	\draw[thin] (0,0,0) -- (\Xthree,0,0);
	\draw[thin] (0,0,\Z) -- (\Xthree,0,\Z);
	\draw[thin, dashed] (\Xone,\Yone,0) -- (\Xtwo,\Yone,0);
	\draw[thin] (\Xone,\Yone,\Z) -- (\Xtwo,\Yone,\Z);
	\draw[thin, dashed] (0,\Ytwo,0) -- (\Xone,\Ytwo,0);
	\draw[thin] (0,\Ytwo,\Z) -- (\Xone,\Ytwo,\Z);
	\draw[thin, dashed] (\Xtwo,\Ytwo,0) -- (\Xthree,\Ytwo,0);
	\draw[thin] (\Xtwo,\Ytwo,\Z) -- (\Xthree,\Ytwo,\Z);
	\draw[thin] (0,0,0) -- (0,\Ytwo,0);
	\draw[thin, dashed] (\Xthree,0,0) -- (\Xthree,\Ytwo,0);
	\draw[thin] (0,0,\Z) -- (0,\Ytwo,\Z);
	\draw[thin] (\Xthree,0,\Z) -- (\Xthree,\Ytwo,\Z);
	\draw[thin, dashed] (\Xtwo,\Yone,0) -- (\Xtwo,\Ytwo,0);
	\draw[thin] (\Xtwo,\Yone,\Z) -- (\Xtwo,\Ytwo,\Z);
	\draw[thin, dashed] (\Xone,\Yone,0) -- (\Xone,\Ytwo,0);
	\draw[thin] (\Xone,\Yone,\Z) -- (\Xone,\Ytwo,\Z);
	\draw[thin] (0,0,0) -- (0,0,\Z);
	\draw[thin] (\Xthree,0,0) -- (\Xthree,0,\Z);
	\draw[thin, dashed] (\Xone,\Yone,0) -- (\Xone,\Yone,\Z);
	\draw[thin, dashed] (\Xtwo,\Yone,0) -- (\Xtwo,\Yone,\Z);
	\draw[thin] (0,\Ytwo,0) -- (0,\Ytwo,\Z);
	\draw[thin, dashed] (\Xone,\Ytwo,0) -- (\Xone,\Ytwo,\Z);
	\draw[thin, dashed] (\Xthree,\Ytwo,0) -- (\Xthree,\Ytwo,\Z);

    \path [overlay, name path=hor_edge] (\Xone,\Yone,\Z) -- (\Xtwo,\Yone,\Z);
    \path [overlay, name path=ver_edge] (\Xtwo,\Ytwo,0) -- (\Xtwo,\Ytwo,\Z);
    \path [overlay, name intersections={of=hor_edge and ver_edge, by=inters_point}];
	\draw[thin, dashed] (\Xtwo,\Ytwo,0) -- (inters_point);
	\draw[thin] (inters_point) -- (\Xtwo,\Ytwo,\Z);
}

\newcommand{\simpleaxes}{%
    \draw[->, very thick] (-0.25,0,0) -- (1,0,0) node[pos=0.75, below]{$x$};
    \draw[->, very thick] (0,-0.25,0) -- (0,1,0) node[pos=0.8, below left]{$y$};
    \draw[->, very thick] (0,0,-0.25) -- (0,0,1) node[pos=0.9, below left]{$z$};
}

\newcommand{\roomsLegend}{%
    \definecolor{myOrange}{RGB}{230, 159, 0}; 
    \definecolor{myLightBlue}{RGB}{86, 180, 233}; 
    \definecolor{myGreen}{RGB}{0, 158, 115}; 
    \definecolor{myYellow}{RGB}{240, 228, 66}; 
    \definecolor{myDarkBlue}{RGB}{0, 114, 178}; 
    \definecolor{myRed}{RGB}{213, 94, 0}; 
    \definecolor{myPurple}{RGB}{204, 121, 167}; 
    
    \draw[pattern={horizontal lines}, pattern color=myGreen] (0.0, 0.0) rectangle +(0.5, 0.5);
    \node at (1.0, 0.25) (input) {$0.95$};
    \draw[fill=gray] (2.0, 0.0) rectangle +(0.5, 0.5);
    \node at (3.0, 0.25) (input) {$0.9$};
    \draw[pattern={vertical lines}, pattern color=myLightBlue] (4.0, 0.0) rectangle +(0.5, 0.5);
    \node at (5.0, 0.25) (input) {$0.8$};
    \draw[pattern={north west lines}, pattern color=myOrange] (6.0, 0.0) rectangle +(0.5, 0.5);
    \node at (7.0, 0.25) (input) {$0.5$};
}

\newif\ifdrawdimlineleft
\newif\ifdrawdimlineright

\pgfkeys{/tikz/.cd,
  angleshift/.store in=\angleshift,
  angleshift=0
}
\tikzset{%
  dimlabel distance/.initial=5mm,
  vertical lines extend/.initial=5mm,
  vertical dim line/.style={gray, thin},
  dim arrow line/.style={%
    arrows={latex[slant={-sin(\angleshift)}]-latex[slant={-sin(\angleshift)}]},
    thin
  },
  dim label/.style={},
  left dimline/.is if=drawdimlineleft,
  left dimline=true,
  right dimline/.is if=drawdimlineright,
  right dimline=true,
  indicate dimensions/.style={%
    decorate, decoration={%
      show path construction,
      lineto code={%
        \draw[dim arrow line]
        ($(\tikzinputsegmentfirst)!\pgfkeysvalueof{/tikz/dimlabel distance}!
        {-90+\angleshift}:(\tikzinputsegmentlast) $)
        -- ($ (\tikzinputsegmentlast)!\pgfkeysvalueof{/tikz/dimlabel distance}!
        {90+\angleshift}:(\tikzinputsegmentfirst) $)
        \ifx#1\empty
        \else
        node[midway, fill=white, dim label,
        sloped, xslant={-sin(\angleshift)}, scale=.8]{#1}
        \fi;
        \ifdrawdimlineleft
        \draw[vertical dim line] (\tikzinputsegmentfirst) -- 
        ($ (\tikzinputsegmentfirst)!\pgfkeysvalueof{/tikz/vertical lines
          extend}!{-90+\angleshift}:(\tikzinputsegmentlast) $);
        \fi
        \ifdrawdimlineright
        \draw[vertical dim line] (\tikzinputsegmentlast) -- 
        ($ (\tikzinputsegmentlast)!\pgfkeysvalueof{/tikz/vertical
          lines extend}!{90+\angleshift}:(\tikzinputsegmentfirst) $);
        \fi 
      }
    }
  }
}

%% file: figures/room_hallway.tex
\begin{tikzpicture}[3d view]
    \shoebox{2}{6}{2}{gray}{gray}{gray!50!white}{gray!50!white}{gray!70!white}{gray!70!white}
    \simpleaxes

    \path [angleshift=-8.5, black, indicate dimensions={$2$}, inner sep=1pt] (2,6,2) -- (0,6,2);
    
    \path [angleshift=24, black, indicate dimensions={$6$}, inner sep=1pt] (2,0,2) -- (2,6,2);

    \path [angleshift=8.5, black, indicate dimensions={\rotatebox[origin=c]{180}{$2$}}, inner sep=1pt] (2,0,0) -- (2,0,2);
    
    \fill (1.2,5.4,1.2) circle (2pt);
    \fill[black,font=\footnotesize] (1.2,5.4,1.2) node [below] {$x_s$};
    
    \fill (0.7,0.6,0.7) circle (2pt);
    \fill[black,font=\footnotesize] (0.7,0.6,0.7) node [above] {$x_r$};
\end{tikzpicture}

%% file: figures/room_uneven.tex
\begin{tikzpicture}[3d view]
    \definecolor{myOrange}{RGB}{230, 159, 0}; 
    \definecolor{myLightBlue}{RGB}{86, 180, 233}; 
    \definecolor{myGreen}{RGB}{0, 158, 115}; 
    \definecolor{myYellow}{RGB}{240, 228, 66}; 
    \definecolor{myDarkBlue}{RGB}{0, 114, 178}; 
    \definecolor{myRed}{RGB}{213, 94, 0}; 
    \definecolor{myPurple}{RGB}{204, 121, 167}; 
    
    \shoebox{5}{6}{3}%
    {pattern={horizontal lines}, pattern color=myGreen}%
    {pattern={horizontal lines}, pattern color=myGreen}%
    {pattern={vertical lines}, pattern color=myLightBlue}%
    {pattern={vertical lines}, pattern color=myLightBlue}%
    {pattern={north west lines}, pattern color=myOrange}%
    {pattern={north west lines}, pattern color=myOrange}
    
    \simpleaxes

    \path [angleshift=-8.5, black, indicate dimensions={$5$}, inner sep=1pt] (5,6,3) -- (0,6,3);
    
    \path [angleshift=24, black, indicate dimensions={$6$}, inner sep=1pt] (5,0,3) -- (5,6,3);

    \path [angleshift=8.5, black, indicate dimensions={\rotatebox[origin=c]{180}{$3$}}, inner sep=1pt] (5,0,0) -- (5,0,3);
    
    \fill (1.2,1.4,1.0) circle (2pt);
    \fill[black,font=\footnotesize] (1.2,1.4,1.0) node [below] {$x_s$};
    
    \fill (0.7,1.6,1.7) circle (2pt);
    \fill[black,font=\footnotesize] (0.7,1.6,1.7) node [above] {$x_r$};
\end{tikzpicture}

%% file: figures/room_non-convex.tex
\begin{tikzpicture}[3d view]
    \nonconvexC{1}{2}{3}{2}{2}
    \simpleaxes

    \path [angleshift=-8.5, black, indicate dimensions={$1$}, inner sep=1pt] (1,5,2) -- (0,5,2);
    \path [angleshift=-8.5, black, indicate dimensions={$2$}, inner sep=1pt] (3,5,2) -- (1,5,2);
    \path [angleshift=-8.5, black, indicate dimensions={$1$}, inner sep=1pt] (4,5,2) -- (3,5,2);
    
    \path [angleshift=24, black, indicate dimensions={$2$}, inner sep=1pt] (4,3,2) -- (4,5,2);
    \path [angleshift=24, black, indicate dimensions={$3$}, inner sep=1pt] (4,0,2) -- (4,3,2);

    \path [angleshift=8.5, black, indicate dimensions={\rotatebox[origin=c]{180}{$2$}}, inner sep=1pt] (4,0,0) -- (4,0,2);
    
    \fill (0.5,4.5,1.0) circle (2pt);
    \fill[black,font=\footnotesize] (0.5,4.5,1.0) node [left] {$x_s$};
    
    \fill (3.5,4.3,1.2) circle (2pt);
    \fill[black,font=\footnotesize] (3.5,4.3,1.2) node [left] {$x_r$};
\end{tikzpicture}

%% file: figures/T30_5.tex
\begin{tikzpicture}
    \definecolor{myOrange}{RGB}{230, 159, 0}; 
    \definecolor{myLightBlue}{RGB}{86, 180, 233}; 
    \definecolor{myGreen}{RGB}{0, 158, 115}; 
    \definecolor{myYellow}{RGB}{240, 228, 66}; 
    \definecolor{myDarkBlue}{RGB}{0, 114, 178}; 
    \definecolor{myRed}{RGB}{213, 94, 0}; 
    \definecolor{myPurple}{RGB}{204, 121, 167}; 
    
    \begin{axis}[
    width=10cm,
    height=6cm,
    scale only axis,
    legend cell align={left},
    legend columns=3,
    legend style={
        at={(0.5, -0.2)},
        anchor=north,
        fill opacity=0.7,
        draw opacity=1,
        text opacity=1,
        draw=lightgray,
        cells={align=left}
    },
    tick align=outside,
    tick pos=left,
    x grid style={darkgray},
    xmode=log,
    xlabel={Frequency [Hz]},
    xmin=125, xmax=16000,
    xtick style={color=black},
    xtick={125,250,500,1000,2000,4000,8000,16000},
    xticklabels={125,250,500,1k,2k,4k,8k,16k},
    y grid style={darkgray},
    ylabel={$T_{30}$ [s]},
    ymin=0.25, ymax=1.05,
    ytick style={color=black},
    mark options={solid}
    ]
    
    \addplot [thick, myLightBlue, mark=*, forget plot]
    table {%
    31.25 0.975002527236938
    62.5 0.981751918792725
    125 0.960474133491516
    250 0.954836845397949
    500 0.981742024421692
    1000 0.951512336730957
    2000 0.901239037513733
    4000 0.80358874797821
    8000 0.630837678909302
    16000 0.476618766784668
    };
    \addplot [thick, myGreen, mark=*, forget plot]
    table {%
    31.25 0.403314919222169
    62.5 0.359095049574131
    125 0.381941845021505
    250 0.405231362372753
    500 0.413051115513987
    1000 0.40558504338691
    2000 0.39471578804448
    4000 0.377012379028655
    8000 0.335820132329085
    16000 0.272121271332936
    };
    \addplot [thick, myOrange, mark=*, forget plot]
    table {%
    31.25 0.48429777000057
    62.5 0.566905867753652
    125 0.569082440406744
    250 0.540066796132852
    500 0.552740880016626
    1000 0.544528084238437
    2000 0.52920753138276
    4000 0.502731664594993
    8000 0.418821618561986
    16000 0.3233157037946
    };
    \addplot [thick, myLightBlue, dashed, mark=square*, forget plot]
    table {%
    31.25 1.07093092127897
    62.5 0.849671196520548
    125 0.949200615473954
    250 0.941024508173289
    500 0.927417993827764
    1000 0.885398028796842
    2000 0.832433094608991
    4000 0.748514579496482
    8000 0.590157898131735
    16000 0.443604139861519
    };
    \addplot [thick, myGreen, dashed, mark=square*, forget plot]
    table {%
    31.25 0.271416346870725
    62.5 0.375897584292593
    125 0.371002804240552
    250 0.395045205672842
    500 0.386189664790171
    1000 0.389057803088746
    2000 0.378212327342299
    4000 0.358299790299343
    8000 0.321918629027563
    16000 0.260221495790848
    };
    \addplot [thick, myOrange, dashed, mark=square*, forget plot]
    table {%
    31.25 0.470765725942516
    62.5 0.566961598853715
    125 0.599613251389404
    250 0.607698484859444
    500 0.605172605056569
    1000 0.59163086376701
    2000 0.562562356785877
    4000 0.520813018150017
    8000 0.438836641439785
    16000 0.33194598778018
    };
    \addplot [thick, myLightBlue, dotted, mark=triangle*, forget plot]
    table {%
    31.25 0.783331413255838
    62.5 0.677072555376966
    125 0.710596362107283
    250 0.744189098554801
    500 0.748059212379831
    1000 0.702356223200468
    2000 0.671564325731775
    4000 0.618578901566346
    8000 0.503269527594304
    16000 0.40037627460508
    };
    \addplot [thick, myGreen, dotted, mark=triangle*, forget plot]
    table {%
    31.25 0.518455652469112
    62.5 0.455269880639259
    125 0.47445489433936
    250 0.487995802785948
    500 0.478379265247278
    1000 0.463883524252434
    2000 0.457026735915546
    4000 0.434645351011683
    8000 0.382769148163682
    16000 0.319861080493886
    };
    \addplot [thick, myOrange, dotted, mark=triangle*, forget plot]
    table {%
    31.25 0.577016542477435
    62.5 0.582586495535118
    125 0.600341970369661
    250 0.630036571389921
    500 0.667680235206744
    1000 0.574374328881847
    2000 0.551326750062974
    4000 0.521143381630539
    8000 0.432442720786225
    16000 0.343606371856243
    };
    \addplot [thick, myPurple, mark=diamond*, mark size=2.5pt, forget plot]
    table {%
    31.25 0.67294431765673
    62.5 0.70343061055884
    125 0.725111344519567
    250 0.527737974779202
    500 0.552399547111004
    1000 0.57789638970998
    2000 0.488148383777305
    4000 0.423548121334537
    8000 0.391726473115386
    16000 0.365976392284542
    };
    \addplot [thick, black, mark=diamond*, mark size=2.5pt, forget plot]
    table {%
    31.25 0.257258475714801
    62.5 0.573097535172086
    125 0.680851005897053
    250 0.579573412775348
    500 0.591299412940376
    1000 0.629245521598483
    2000 0.59716731030225
    4000 0.533894867029331
    8000 0.460231617743905
    16000 0.390570258509883
    };
    
    \addplot [line width=1.5pt, myPurple, mark=diamond*, mark size=3.5pt]
    table {%
    1 -1
    2 -1
    };
    \addlegendentry{PyRoomAcoustics}
    \addplot [line width=1.5pt, black, mark=diamond*, mark size=3.5pt]
    table {%
    1 -1
    2 -1
    };
    \addlegendentry{CATT-Acoustics}
    \addlegendimage{empty legend}\addlegendentry{}
    \addplot [line width=1.5pt, myLightBlue, mark=*, mark size=3.5pt]
    table {%
    1 -1
    2 -1
    };
    \addlegendentry{ARN ${M=3328}$\\(baseline)}
    \addplot [line width=1.5pt, myGreen, mark=*, mark size=3.5pt]
    table {%
    1 -1
    2 -1
    };
    \addlegendentry{ARN ${M=3328}$\\(Sinkhorn-Knopp)}
    \addplot [line width=1.5pt, myOrange, mark=*, mark size=3.5pt]
    table {%
    1 -1
    2 -1
    };
    \addlegendentry{ARN ${M=3328}$\\(uniform)}
    \addplot [line width=1.5pt, myLightBlue, dashed, mark=square*, mark size=3.5pt]
    table {%
    1 -1
    2 -1
    };
    \addlegendentry{ARN ${M=1458}$\\(baseline)}
    \addplot [line width=1.5pt, myGreen, dashed, mark=square*, mark size=3.5pt]
    table {%
    1 -1
    2 -1
    };
    \addlegendentry{ARN ${M=1458}$\\(Sinkhorn-Knopp)}
    \addplot [line width=1.5pt, myOrange, dashed, mark=square*, mark size=3.5pt]
    table {%
    1 -1
    2 -1
    };
    \addlegendentry{ARN ${M=1458}$\\(uniform)}
    \addplot [line width=1.5pt, myLightBlue, dotted, mark=triangle*, mark size=3.5pt]
    table {%
    1 -1
    2 -1
    };
    \addlegendentry{ARN ${M=208}$\\(baseline)}
    \addplot [line width=1.5pt, myGreen, dotted, mark=triangle*, mark size=3.5pt]
    table {%
    1 -1
    2 -1
    };
    \addlegendentry{ARN ${M=208}$\\(Sinkhorn-Knopp)}
    \addplot [line width=1.5pt, myOrange, dotted, mark=triangle*, mark size=3.5pt]
    table {%
    1 -1
    2 -1
    };
    \addlegendentry{ARN ${M=208}$\\(uniform)}
    
    \end{axis}

\end{tikzpicture}

%% file: figures/T30_4.tex
\begin{tikzpicture}
    \definecolor{myOrange}{RGB}{230, 159, 0}; 
    \definecolor{myLightBlue}{RGB}{86, 180, 233}; 
    \definecolor{myGreen}{RGB}{0, 158, 115}; 
    \definecolor{myYellow}{RGB}{240, 228, 66}; 
    \definecolor{myDarkBlue}{RGB}{0, 114, 178}; 
    \definecolor{myRed}{RGB}{213, 94, 0}; 
    \definecolor{myPurple}{RGB}{204, 121, 167}; 
    
    \begin{axis}[
    width=10cm,
    height=6cm,
    scale only axis,
    legend cell align={left},
    legend columns=3,
    legend style={
        at={(0.02, 0.02)},
        anchor=south west,
        fill opacity=0.7,
        draw opacity=1,
        text opacity=1,
        draw=lightgray,
        cells={align=left},
        font=\footnotesize
    },
    tick align=outside,
    tick pos=left,
    x grid style={darkgray},
    xmode=log,
    xlabel={Frequency [Hz]},
    xmin=125, xmax=16000,
    xtick style={color=black},
    xtick={125,250,500,1000,2000,4000,8000,16000},
    xticklabels={125,250,500,1k,2k,4k,8k,16k},
    y grid style={darkgray},
    ylabel={$T_{30}$ [s]},
    ymin=0.35, ymax=0.85,
    ytick style={color=black},
    mark options={solid}
    ]
    
    \addplot [thick, myLightBlue, mark=*, forget plot]
    table {%
    31.25 0.671785593032837
    62.5 0.666484832763672
    125 0.754998683929443
    250 0.825580358505249
    500 0.808791875839233
    1000 0.813389182090759
    2000 0.79275918006897
    4000 0.748466610908508
    8000 0.661300897598267
    16000 0.602038860321045
    };
    \addplot [thick, myGreen, mark=*, forget plot]
    table {%
    31.25 0.753826899716313
    62.5 0.677234433222509
    125 0.697765094680693
    250 0.678036129676397
    500 0.655010995648017
    1000 0.65322423314273
    2000 0.631088589218245
    4000 0.591860397870995
    8000 0.516854271769078
    16000 0.458680200171988
    };
    \addplot [thick, myOrange, mark=*, forget plot]
    table {%
    31.25 0.701982386382271
    62.5 0.723519722907846
    125 0.660458773534352
    250 0.648832861005267
    500 0.670426762101497
    1000 0.646870599669677
    2000 0.62303921652979
    4000 0.583392168782384
    8000 0.519037112715406
    16000 0.460031438142471
    };
    \addplot [thick, myLightBlue, dashed, mark=square*, forget plot]
    table {%
    31.25 0.677138396872244
    62.5 0.736111512353202
    125 0.730494958798273
    250 0.707970026748693
    500 0.732109662281925
    1000 0.730458062367091
    2000 0.690061532540117
    4000 0.660184845379646
    8000 0.582441029834276
    16000 0.543513801846782
    };
    \addplot [thick, myGreen, dashed, mark=square*, forget plot]
    table {%
    31.25 0.718108428574058
    62.5 0.706595762197832
    125 0.717198164886305
    250 0.681306588803835
    500 0.665048363659012
    1000 0.650371442742398
    2000 0.644093699028835
    4000 0.605061234682484
    8000 0.530843973194939
    16000 0.473564141758991
    };
    \addplot [thick, myOrange, dashed, mark=square*, forget plot]
    table {%
    31.25 0.635688688354734
    62.5 0.66489298788576
    125 0.704925193901409
    250 0.682209088402263
    500 0.655859588815458
    1000 0.657581990377539
    2000 0.645432828851568
    4000 0.607395641458697
    8000 0.537025898967513
    16000 0.478775483023881
    };
    \addplot [thick, myLightBlue, dotted, mark=triangle*, forget plot]
    table {%
    31.25 0.73738411247694
    62.5 0.682543194185126
    125 0.660081667854446
    250 0.646725395441142
    500 0.643193879952408
    1000 0.662466928428231
    2000 0.65892512679611
    4000 0.611824168983982
    8000 0.539802537821173
    16000 0.487342050068215
    };
    \addplot [thick, myGreen, dotted, mark=triangle*, forget plot]
    table {%
    31.25 0.621508935383367
    62.5 0.632668006307044
    125 0.625268287773373
    250 0.638028234669379
    500 0.651725639734133
    1000 0.63572853279076
    2000 0.617613980852388
    4000 0.577700711210093
    8000 0.514562645290968
    16000 0.454188767243431
    };
    \addplot [thick, myOrange, dotted, mark=triangle*, forget plot]
    table {%
    31.25 0.712940085035806
    62.5 0.664150195989872
    125 0.682073358331422
    250 0.660080057705933
    500 0.636510482715127
    1000 0.648958819781832
    2000 0.62389831905732
    4000 0.590213669547005
    8000 0.52275619347521
    16000 0.462411303321472
    };
    \addplot [thick, black, mark=diamond*, mark size=2.5pt, forget plot]
    table {%
    31.25 0.523689844568521
    62.5 0.579846714504046
    125 0.654419765672999
    250 0.651401014291232
    500 0.636879226469063
    1000 0.641446290289316
    2000 0.618923355035378
    4000 0.585031409182431
    8000 0.517400644050491
    16000 0.4821927803059
    };
    
    \addplot [line width=1.25pt, black, mark=diamond*, mark size=3pt]
    table {%
    1 -1
    2 -1
    };
    \addlegendentry{CATT-Acoustics}
    \addlegendimage{empty legend}\addlegendentry{}
    \addlegendimage{empty legend}\addlegendentry{}
    \addplot [line width=1.25pt, myLightBlue, mark=*, mark size=3pt]
    table {%
    1 -1
    2 -1
    };
    \addlegendentry{ARN ${M=1954}$\\(baseline)}
    \addplot [line width=1.25pt, myGreen, mark=*, mark size=3pt]
    table {%
    1 -1
    2 -1
    };
    \addlegendentry{ARN ${M=1954}$\\(Sinkhorn-Knopp)}
    \addplot [line width=1.25pt, myOrange, mark=*, mark size=3pt]
    table {%
    1 -1
    2 -1
    };
    \addlegendentry{ARN ${M=1954}$\\(uniform)}
    \addplot [line width=1.25pt, myLightBlue, dashed, mark=square*, mark size=3pt]
    table {%
    1 -1
    2 -1
    };
    \addlegendentry{ARN ${M=232}$\\(baseline)}
    \addplot [line width=1.25pt, myGreen, dashed, mark=square*, mark size=3pt]
    table {%
    1 -1
    2 -1
    };
    \addlegendentry{ARN ${M=232}$\\(Sinkhorn-Knopp)}
    \addplot [line width=1.25pt, myOrange, dashed, mark=square*, mark size=3pt]
    table {%
    1 -1
    2 -1
    };
    \addlegendentry{ARN ${M=232}$\\(uniform)}
    \addplot [line width=1.25pt, myLightBlue, dotted, mark=triangle*, mark size=3pt]
    table {%
    1 -1
    2 -1
    };
    \addlegendentry{ARN ${M=118}$\\(baseline)}
    \addplot [line width=1.25pt, myGreen, dotted, mark=triangle*, mark size=3pt]
    table {%
    1 -1
    2 -1
    };
    \addlegendentry{ARN ${M=118}$\\(Sinkhorn-Knopp)}
    \addplot [line width=1.25pt, myOrange, dotted, mark=triangle*, mark size=3pt]
    table {%
    1 -1
    2 -1
    };
    \addlegendentry{ARN ${M=118}$\\(uniform)}
    
    \end{axis}
    
\end{tikzpicture}

%% file: figures/EDT_2.tex
\begin{tikzpicture}
    \definecolor{myOrange}{RGB}{230, 159, 0}; 
    \definecolor{myLightBlue}{RGB}{86, 180, 233}; 
    \definecolor{myGreen}{RGB}{0, 158, 115}; 
    \definecolor{myYellow}{RGB}{240, 228, 66}; 
    \definecolor{myDarkBlue}{RGB}{0, 114, 178}; 
    \definecolor{myRed}{RGB}{213, 94, 0}; 
    \definecolor{myPurple}{RGB}{204, 121, 167}; 
    
    \begin{axis}[
    width=10cm,
    height=6cm,
    scale only axis,
    legend cell align={left},
    legend columns=2,
    legend style={
        at={(0.02, 0.02)},
        anchor=south west,
        fill opacity=0.7,
        draw opacity=1,
        text opacity=1,
        draw=lightgray
    },
    tick align=outside,
    tick pos=left,
    x grid style={darkgray},
    xmode=log,
    xlabel={Frequency [Hz]},
    xmin=125, xmax=16000,
    xtick style={color=black},
    xtick={125,250,500,1000,2000,4000,8000,16000},
    xticklabels={125,250,500,1k,2k,4k,8k,16k},
    y grid style={darkgray},
    ylabel={EDT [ms]},
    ymin=65, ymax=185,
    ytick style={color=black}
    ]
    
    \addplot [thick, myDarkBlue, mark=x, forget plot, mark size=2.5pt]
    table {%
    31.25 137.672134399414
    62.5 118.348075866699
    125 110.74104309082
    250 108.531478881836
    500 114.636024475098
    1000 113.481964111328
    2000 119.512023925781
    4000 112.222373962402
    8000 94.8671569824219
    16000 75.9612731933594
    };
    \addplot [thick, myGreen, mark=otimes, forget plot, mark size=2.5pt]
    table {%
    31.25 83.5294216473015
    62.5 66.8187485150845
    125 111.925487382162
    250 129.470781720321
    500 125.00457691775
    1000 115.927869774775
    2000 120.81598848756
    4000 106.65688298071
    8000 91.1256826639872
    16000 71.2099972718421
    };
    \addplot [thick, myOrange, mark=+, forget plot, mark size=2.5pt]
    table {%
    31.25 117.908503380998
    62.5 111.467942661766
    125 123.945475779042
    250 110.390790548867
    500 123.853360593958
    1000 110.538906896736
    2000 114.704984514363
    4000 112.169017875266
    8000 98.3546854506973
    16000 78.4885329452345
    };
    \addplot [thick, myRed, mark=oplus, forget plot, mark size=2.5pt]
    table {%
    31.25 136.637264709457
    62.5 128.174907832572
    125 149.406326630874
    250 124.038965597971
    500 127.217847807247
    1000 122.775528857254
    2000 117.071827737234
    4000 107.421779551815
    8000 88.7752052063497
    16000 70.5988234850567
    };
    \addplot [thick, myLightBlue, mark=*, forget plot, mark size=2.5pt]
    table {%
    31.25 105.761677419353
    62.5 78.7454819872436
    125 85.5554010530806
    250 117.257570645131
    500 160.459482031532
    1000 176.888894055461
    2000 171.655759642919
    4000 153.06252379285
    8000 120.90358544438
    16000 87.2191683670204
    };
    \addplot [thick, black, mark=diamond*, forget plot, mark size=2.5pt]
    table {%
    31.25 97.7320495818991
    62.5 86.8973713510993
    125 121.551390416191
    250 131.640655537904
    500 128.359834514866
    1000 126.129001452938
    2000 115.486337290895
    4000 104.165980414291
    8000 98.5527159144505
    16000 84.2752660995743
    };
    
    \addplot [line width=1.25pt, black, mark=diamond*, mark size=4pt]
    table {%
    1 -1
    2 -1
    };
    \addlegendentry{CATT-Acoustics}
    \addplot [line width=1.25pt, myLightBlue, mark=*, mark size=4pt]
    table {%
    1 -1
    2 -1
    };
    \addlegendentry{ARN (baseline)}
    \addplot [line width=1.25pt, myDarkBlue, mark=x, mark size=4pt]
    table {%
    1 -1
    2 -1
    };
    \addlegendentry{ARN (${K=1}$)}
    \addplot [line width=1.25pt, myGreen, mark=otimes, mark size=4pt]
    table {%
    1 -1
    2 -1
    };
    \addlegendentry{ARN (${K=3}$)}
    \addplot [line width=1.25pt, myOrange, mark=+, mark size=4pt]
    table {%
    1 -1
    2 -1
    };
    \addlegendentry{ARN (${K=1}$, spr.)}
    \addplot [line width=1.25pt, myRed, mark=oplus, mark size=4pt]
    table {%
    1 -1
    2 -1
    };
    \addlegendentry{ARN (${K=3}$, spr.)}
    
    \end{axis}

\end{tikzpicture}

%% file: figures/cr2_really_simplified.tex
\begin{tikzpicture}[3d view = {-30}{15}]
	\newcommand \FrontOpacity {0.5};
	\newcommand \BackOpacity {0.9};
	
	\coordinate (V0) at (-1.303, 4.242, 0.15);
	\coordinate (V1) at (-2.823, 4.082, 0.15);
	\coordinate (V2) at (-2.823, 4.242, 0.15);
	\coordinate (V3) at (-1.303, 4.082, 0.15);
	\coordinate (V4) at (-1.303, 4.082, 0.595);
	\coordinate (V5) at (-2.823, 4.242, 0.595);
	\coordinate (V6) at (-2.823, 4.082, 0.595);
	\coordinate (V7) at (-1.303, 4.242, 0.595);
	\coordinate (V8) at (0.757, 4.242, 0.15);
	\coordinate (V9) at (-0.763, 4.082, 0.15);
	\coordinate (V10) at (-0.763, 4.242, 0.15);
	\coordinate (V11) at (0.757, 4.082, 0.15);
	\coordinate (V12) at (0.757, 4.082, 0.595);
	\coordinate (V13) at (-0.763, 4.242, 0.595);
	\coordinate (V14) at (-0.763, 4.082, 0.595);
	\coordinate (V15) at (0.757, 4.242, 0.595);
	\coordinate (V16) at (2.804, 4.337, 2.568);
	\coordinate (V17) at (1.261, 4.337, 2.568);
	\coordinate (V18) at (2.804, 4.337, 0.45);
	\coordinate (V19) at (1.261, 4.337, 0.45);
	\coordinate (V20) at (3.132, -1.695, 0.0);
	\coordinate (V21) at (3.132, 3.815, 0.0);
	\coordinate (V22) at (3.132, 3.815, 2.988);
	\coordinate (V23) at (3.132, -1.695, 2.988);
	\coordinate (V24) at (2.869, 3.815, 0.0);
	\coordinate (V25) at (2.869, 4.189, 2.988);
	\coordinate (V26) at (2.869, 3.815, 2.988);
	\coordinate (V27) at (2.869, 4.189, 0.0);
	\coordinate (V28) at (-2.884, 3.785, 0.0);
	\coordinate (V29) at (-3.036, 3.785, 2.5);
	\coordinate (V30) at (-3.036, 3.785, 0.0);
	\coordinate (V31) at (-2.884, 3.785, 2.5);
	\coordinate (V32) at (-3.036, -2.891, 2.5);
	\coordinate (V33) at (-2.884, -2.891, 2.5);
	\coordinate (V34) at (-3.036, -2.891, 0.0);
	\coordinate (V35) at (-1.616, -2.891, 0.0);
	\coordinate (V36) at (3.094, -4.255, 0.0);
	\coordinate (V37) at (-1.616, -4.255, 0.0);
	\coordinate (V38) at (2.489, -3.695, 0.0);
	\coordinate (V39) at (3.094, -3.695, 0.0);
	\coordinate (V40) at (2.489, -1.695, 0.0);
	\coordinate (V41) at (-2.884, 4.189, 0.0);
	\coordinate (V42) at (3.094, -3.695, 2.988);
	\coordinate (V43) at (3.094, -4.255, 1.111);
	\coordinate (V44) at (3.094, -4.255, 2.988);
	\coordinate (V45) at (3.094, -3.942, 1.111);
	\coordinate (V46) at (3.094, -3.942, 0.907);
	\coordinate (V47) at (3.094, -4.255, 0.578);
	\coordinate (V48) at (2.489, -3.695, 2.988);
	\coordinate (V49) at (2.489, -1.695, 2.988);
	\coordinate (V50) at (-1.616, -4.255, 2.988);
	\coordinate (V51) at (2.902, -4.255, 0.578);
	\coordinate (V52) at (2.902, -4.255, 1.111);
	\coordinate (V53) at (-1.616, -2.881, 2.988);
	\coordinate (V54) at (-2.884, -2.881, 2.988);
	\coordinate (V55) at (-2.884, 4.189, 2.988);
	\coordinate (V56) at (2.902, -3.942, 1.111);
	\coordinate (V57) at (2.902, -3.942, 0.907);
	\coordinate (V58) at (0.758, 4.337, 2.568);
	\coordinate (V59) at (-0.769, 4.337, 2.568);
	\coordinate (V60) at (-2.829, 4.337, 2.568);
	\coordinate (V61) at (-2.829, 4.337, 0.45);
	\coordinate (V62) at (0.758, 4.337, 0.45);
	\coordinate (V63) at (-1.297, 4.337, 2.568);
	\coordinate (V64) at (-1.297, 4.337, 0.45);
	\coordinate (V65) at (-0.769, 4.337, 0.45);
	\coordinate (V66) at (1.267, 4.242, 0.595);
	\coordinate (V67) at (1.267, 4.082, 0.15);
	\coordinate (V68) at (1.267, 4.082, 0.595);
	\coordinate (V69) at (1.267, 4.242, 0.15);
	\coordinate (V70) at (2.787, 4.242, 0.15);
	\coordinate (V71) at (2.787, 4.082, 0.15);
	\coordinate (V72) at (2.787, 4.082, 0.595);
	\coordinate (V73) at (2.787, 4.242, 0.595);
	\coordinate (V74) at (-1.616, -2.891, 2.5);
	\coordinate (V75) at (-1.616, -3.695, 0.0);
	\coordinate (V76) at (2.489, -2.891, 0.0);
	\coordinate (V77) at (-3.036, -1.695, 0.0);
	\coordinate (V78) at (2.869, -1.695, 0.0);
	\coordinate (V79) at (2.869, 4.337, 0.0);
	\coordinate (V80) at (2.869, 4.337, 2.568);
	\coordinate (V81) at (2.869, 4.337, 0.45);
	\coordinate (V82) at (2.869, 4.189, 2.568);
	\coordinate (V83) at (-2.884, 4.189, 2.568);
	\coordinate (V84) at (-2.884, 4.337, 0.0);
	\coordinate (V85) at (-2.884, 4.337, 2.568);
	\coordinate (V86) at (-2.884, 4.337, 0.45);
	\coordinate (V87) at (-2.884, 3.785, 2.988);
	\coordinate (V88) at (-1.616, -3.695, 2.988);
	\coordinate (V89) at (2.869, -1.695, 2.988);
	\coordinate (V90) at (-2.884, 3.815, 2.988);
	\coordinate (V91) at (-2.884, -1.695, 2.988);
	\coordinate (V92) at (2.489, -2.881, 2.988);
 
	\draw[arrows = {->[harpoon, swap, length=5pt, width=6pt]}, thick, transform canvas={yshift=-0.25pt}] (-0.25, 0, 0) -- (1, 0, 0);
	\draw[arrows = {->[harpoon, length=5pt, width=6pt]}, thick, transform canvas={yshift=-0.25pt}] (0, -0.25, 0) -- (0, 1, 0);
	\draw[very thick] (0, 0, -0.25) -- (0, 0, 0);
	
 
	\path[thin, dashed, draw = {rgb,255:red,104; green,255; blue,242}, fill = {rgb,255:red,104; green,255; blue,242}, fill opacity = \BackOpacity] (V60) -- (V63) -- (V64) -- (V61) -- cycle;
	\path[thin, dashed, draw = {rgb,255:red,104; green,255; blue,242}, fill = {rgb,255:red,104; green,255; blue,242}, fill opacity = \BackOpacity] (V59) -- (V58) -- (V62) -- (V65) -- cycle;
	\path[thin, dashed, draw = {rgb,255:red,104; green,255; blue,242}, fill = {rgb,255:red,104; green,255; blue,242}, fill opacity = \BackOpacity] (V17) -- (V16) -- (V18) -- (V19) -- cycle;
 
 
	\path[thin, dashed, draw = {rgb,255:red,178; green,171; blue,137}, fill = {rgb,255:red,178; green,171; blue,137}, fill opacity = \BackOpacity] (V37) -- (V35) -- (V34) -- (V30) -- (V28) -- (V84) -- (V79) -- (V24) -- (V21) -- (V20) -- (V40) -- (V38) -- (V39) -- (V36) -- cycle;

	\draw[arrows = {->[harpoon, length=5pt, width=6pt]}, thick, transform canvas={yshift=0.25pt}] (-0.25, 0, 0) -- (1, 0, 0) node[pos = 0.75, above]{$x$};
	\draw[arrows = {->[harpoon, swap, length=5pt, width=6pt]}, thick, transform canvas={yshift=0.25pt}] (0, -0.25, 0) -- (0, 1, 0) node[pos = 0.9, above]{$y$};
	\draw[arrows = {->[length=5pt, width=6pt]}, very thick] (0, 0, 0) -- (0, 0, 1) node[pos = 0.8, above right]{$z$};
 
	\path[thin, draw = {rgb,255:red,142; green,141; blue,130}, fill = {rgb,255:red,142; green,141; blue,130}, fill opacity = \FrontOpacity] (V50) -- (V53) -- (V35) -- (V37) -- cycle;
	\path[thin, dashed, draw = {rgb,255:red,142; green,141; blue,130}, fill = {rgb,255:red,142; green,141; blue,130}, fill opacity = \BackOpacity] (V36) -- (V39) -- (V42) -- (V44) -- cycle;
	\path[thin, dashed, draw = {rgb,255:red,142; green,141; blue,130}, fill = {rgb,255:red,142; green,141; blue,130}, fill opacity = \BackOpacity] (V39) -- (V38) -- (V48) -- (V42) -- cycle;
	\path[thin, dashed, draw = {rgb,255:red,142; green,141; blue,130}, fill = {rgb,255:red,142; green,141; blue,130}, fill opacity = \BackOpacity] (V49) -- (V48) -- (V38) -- (V40) -- cycle;
	\path[thin, draw = {rgb,255:red,142; green,141; blue,130}, fill = {rgb,255:red,142; green,141; blue,130}, fill opacity = \FrontOpacity] (V40) -- (V20) -- (V23) -- (V49) -- cycle;
	\path[thin, dashed, draw = {rgb,255:red,142; green,141; blue,130}, fill = {rgb,255:red,142; green,141; blue,130}, fill opacity = \BackOpacity] (V20) -- (V21) -- (V22) -- (V23) -- cycle;
	\path[thin, dashed, draw = {rgb,255:red,142; green,141; blue,130}, fill = {rgb,255:red,142; green,141; blue,130}, fill opacity = \BackOpacity] (V21) -- (V24) -- (V26) -- (V22) -- cycle;
	\path[thin, dashed, draw = {rgb,255:red,142; green,141; blue,130}, fill = {rgb,255:red,142; green,141; blue,130}, fill opacity = \BackOpacity] (V24) -- (V27) -- (V25) -- (V26) -- cycle;
	\path[thin, dashed, draw = {rgb,255:red,142; green,141; blue,130}, fill = {rgb,255:red,142; green,141; blue,130}, fill opacity = \BackOpacity] (V27) -- (V79) -- (V80) -- (V82) -- cycle;
	\path[thin, dashed, draw = {rgb,255:red,142; green,141; blue,130}, fill = {rgb,255:red,142; green,141; blue,130}, fill opacity = \BackOpacity] (V82) -- (V83) -- (V55) -- (V25) -- cycle;
	\path[thin, draw = {rgb,255:red,142; green,141; blue,130}, fill = {rgb,255:red,142; green,141; blue,130}, fill opacity = \FrontOpacity] (V82) -- (V80) -- (V85) -- (V83) -- cycle;
	\path[thin, dashed, draw = {rgb,255:red,142; green,141; blue,130}, fill = {rgb,255:red,142; green,141; blue,130}, fill opacity = \BackOpacity] (V79) -- (V84) -- (V86) -- (V81) -- cycle;
	\path[thin, dashed, draw = {rgb,255:red,142; green,141; blue,130}, fill = {rgb,255:red,142; green,141; blue,130}, fill opacity = \BackOpacity] (V81) -- (V18) -- (V16) -- (V80) -- cycle;
	\path[thin, dashed, draw = {rgb,255:red,142; green,141; blue,130}, fill = {rgb,255:red,142; green,141; blue,130}, fill opacity = \BackOpacity] (V61) -- (V86) -- (V85) -- (V60) -- cycle;
	\path[thin, dashed, draw = {rgb,255:red,142; green,141; blue,130}, fill = {rgb,255:red,142; green,141; blue,130}, fill opacity = \BackOpacity] (V19) -- (V62) -- (V58) -- (V17) -- cycle;
	\path[thin, dashed, draw = {rgb,255:red,142; green,141; blue,130}, fill = {rgb,255:red,142; green,141; blue,130}, fill opacity = \BackOpacity] (V65) -- (V64) -- (V63) -- (V59) -- cycle;
	\path[thin, draw = {rgb,255:red,142; green,141; blue,130}, fill = {rgb,255:red,142; green,141; blue,130}, fill opacity = \FrontOpacity] (V84) -- (V41) -- (V83) -- (V85) -- cycle;
	\path[thin, draw = {rgb,255:red,142; green,141; blue,130}, fill = {rgb,255:red,142; green,141; blue,130}, fill opacity = \FrontOpacity] (V41) -- (V28) -- (V87) -- (V55) -- cycle;
	\path[thin, dashed, draw = {rgb,255:red,142; green,141; blue,130}, fill = {rgb,255:red,142; green,141; blue,130}, fill opacity = \BackOpacity] (V28) -- (V30) -- (V29) -- (V31) -- cycle;
	\path[thin, draw = {rgb,255:red,142; green,141; blue,130}, fill = {rgb,255:red,142; green,141; blue,130}, fill opacity = \FrontOpacity] (V31) -- (V33) -- (V54) -- (V87) -- cycle;
	\path[thin, draw = {rgb,255:red,142; green,141; blue,130}, fill = {rgb,255:red,142; green,141; blue,130}, fill opacity = \FrontOpacity] (V31) -- (V29) -- (V32) -- (V33) -- cycle;
	\path[thin, draw = {rgb,255:red,142; green,141; blue,130}, fill = {rgb,255:red,142; green,141; blue,130}, fill opacity = \FrontOpacity] (V71) -- (V72) -- (V73) -- (V70) -- cycle;
	\path[thin, dashed, draw = {rgb,255:red,142; green,141; blue,130}, fill = {rgb,255:red,142; green,141; blue,130}, fill opacity = \BackOpacity] (V71) -- (V67) -- (V68) -- (V72) -- cycle;
	\path[thin, dashed, draw = {rgb,255:red,142; green,141; blue,130}, fill = {rgb,255:red,142; green,141; blue,130}, fill opacity = \BackOpacity] (V68) -- (V67) -- (V69) -- (V66) -- cycle;
	\path[thin, draw = {rgb,255:red,142; green,141; blue,130}, fill = {rgb,255:red,142; green,141; blue,130}, fill opacity = \FrontOpacity] (V66) -- (V69) -- (V70) -- (V73) -- cycle;
	\path[thin, dashed, draw = {rgb,255:red,142; green,141; blue,130}, fill = {rgb,255:red,142; green,141; blue,130}, fill opacity = \BackOpacity] (V72) -- (V68) -- (V66) -- (V73) -- cycle;
	\path[thin, draw = {rgb,255:red,142; green,141; blue,130}, fill = {rgb,255:red,142; green,141; blue,130}, fill opacity = \FrontOpacity] (V71) -- (V70) -- (V69) -- (V67) -- cycle;
	\path[thin, draw = {rgb,255:red,142; green,141; blue,130}, fill = {rgb,255:red,142; green,141; blue,130}, fill opacity = \FrontOpacity] (V11) -- (V12) -- (V15) -- (V8) -- cycle;
	\path[thin, dashed, draw = {rgb,255:red,142; green,141; blue,130}, fill = {rgb,255:red,142; green,141; blue,130}, fill opacity = \BackOpacity] (V9) -- (V14) -- (V12) -- (V11) -- cycle;
	\path[thin, dashed, draw = {rgb,255:red,142; green,141; blue,130}, fill = {rgb,255:red,142; green,141; blue,130}, fill opacity = \BackOpacity] (V9) -- (V10) -- (V13) -- (V14) -- cycle;
	\path[thin, draw = {rgb,255:red,142; green,141; blue,130}, fill = {rgb,255:red,142; green,141; blue,130}, fill opacity = \FrontOpacity] (V13) -- (V10) -- (V8) -- (V15) -- cycle;
	\path[thin, dashed, draw = {rgb,255:red,142; green,141; blue,130}, fill = {rgb,255:red,142; green,141; blue,130}, fill opacity = \BackOpacity] (V12) -- (V14) -- (V13) -- (V15) -- cycle;
	\path[thin, draw = {rgb,255:red,142; green,141; blue,130}, fill = {rgb,255:red,142; green,141; blue,130}, fill opacity = \FrontOpacity] (V11) -- (V8) -- (V10) -- (V9) -- cycle;
	\path[thin, draw = {rgb,255:red,142; green,141; blue,130}, fill = {rgb,255:red,142; green,141; blue,130}, fill opacity = \FrontOpacity] (V0) -- (V3) -- (V4) -- (V7) -- cycle;
	\path[thin, dashed, draw = {rgb,255:red,142; green,141; blue,130}, fill = {rgb,255:red,142; green,141; blue,130}, fill opacity = \BackOpacity] (V3) -- (V1) -- (V6) -- (V4) -- cycle;
	\path[thin, dashed, draw = {rgb,255:red,142; green,141; blue,130}, fill = {rgb,255:red,142; green,141; blue,130}, fill opacity = \BackOpacity] (V1) -- (V2) -- (V5) -- (V6) -- cycle;
	\path[thin, draw = {rgb,255:red,142; green,141; blue,130}, fill = {rgb,255:red,142; green,141; blue,130}, fill opacity = \FrontOpacity] (V5) -- (V2) -- (V0) -- (V7) -- cycle;
	\path[thin, dashed, draw = {rgb,255:red,142; green,141; blue,130}, fill = {rgb,255:red,142; green,141; blue,130}, fill opacity = \BackOpacity] (V4) -- (V6) -- (V5) -- (V7) -- cycle;
	\path[thin, draw = {rgb,255:red,142; green,141; blue,130}, fill = {rgb,255:red,142; green,141; blue,130}, fill opacity = \FrontOpacity] (V3) -- (V0) -- (V2) -- (V1) -- cycle;
 
	\path[thin, draw = {rgb,255:red,216; green,214; blue,198}, fill = {rgb,255:red,216; green,214; blue,198}, fill opacity = \FrontOpacity] (V29) -- (V30) -- (V34) -- (V32) -- cycle;
	\path[thin, draw = {rgb,255:red,216; green,214; blue,198}, fill = {rgb,255:red,216; green,214; blue,198}, fill opacity = \FrontOpacity] (V34) -- (V35) -- (V53) -- (V54) -- (V33) -- (V32) -- cycle;
	\path[thin, draw = {rgb,255:red,216; green,214; blue,198}, fill = {rgb,255:red,216; green,214; blue,198}, fill opacity = \FrontOpacity] (V37) -- (V36) -- (V44) -- (V50) -- cycle;
 

    \path[thin, draw = {rgb,255:red,226; green,226; blue,226}, fill = {rgb,255:red,226; green,226; blue,226}, fill opacity = \FrontOpacity] (V44) -- (V42) -- (V48) -- (V49) -- (V23) -- (V22) -- (V26) -- (V25) -- (V55) -- (V54) -- (V53) -- (V88) -- (V50) -- cycle;
\end{tikzpicture}

%% file: figures/T30_BRAS.tex
\begin{tikzpicture}
    \definecolor{myOrange}{RGB}{230, 159, 0}; 
    \definecolor{myLightBlue}{RGB}{86, 180, 233}; 
    \definecolor{myGreen}{RGB}{0, 158, 115}; 
    \definecolor{myYellow}{RGB}{240, 228, 66}; 
    \definecolor{myDarkBlue}{RGB}{0, 114, 178}; 
    \definecolor{myRed}{RGB}{213, 94, 0}; 
    \definecolor{myPurple}{RGB}{204, 121, 167}; 
    
    \begin{axis}[
    width=10cm,
    height=6cm,
    scale only axis,
    legend cell align={left},
    legend columns=2,
    legend style={
        at={(0.02, 0.02)},
        anchor=south west,
        fill opacity=0.7,
        draw opacity=1,
        text opacity=1,
        draw=lightgray
    },
    tick align=outside,
    tick pos=left,
    x grid style={darkgray},
    xmode=log,
    xlabel={Frequency [Hz]},
    xmin=125, xmax=16000,
    xtick style={color=black},
    xtick={125,250,500,1000,2000,4000,8000,16000},
    xticklabels={125,250,500,1k,2k,4k,8k,16k},
    y grid style={darkgray},
    ylabel={$T_{30}$ [s]},
    ymin=0.75, ymax=2.05,
    ytick style={color=black},
    mark options={solid}
    ]
    
    \addplot [thick, myLightBlue, mark=*, forget plot, mark size=2.5pt]
    table {%
    125 1.40072441101074
    249.999954223633 1.73005771636963
    500.000091552734 1.96304535865784
    1000 1.85260319709778
    1999.99963378906 1.68329739570618
    4000.00073242188 1.51692616939545
    8000 1.17765808105469
    16000.0068359375 0.80850076675415
    };
    \addplot [thick, myDarkBlue, mark=x, forget plot, mark size=2.5pt]
    table {%
    125 1.52028667926788
    249.999954223633 1.7894561290741
    500.000091552734 1.91489279270172
    1000 1.83303463459015
    1999.99963378906 1.66874563694
    4000.00073242188 1.50271570682526
    8000 1.18335902690887
    16000.0068359375 0.821804523468018
    };
    \addplot [thick, myGreen, mark=otimes, forget plot, mark size=2.5pt]
    table {%
    125 1.44418776035309
    249.999954223633 1.77056050300598
    500.000091552734 1.88174140453339
    1000 1.81798493862152
    1999.99963378906 1.68115830421448
    4000.00073242188 1.5250461101532
    8000 1.19067168235779
    16000.0068359375 0.778941750526428
    };
    \addplot [thick, myOrange, mark=+, forget plot, mark size=2.5pt]
    table {%
    125 1.55929183959961
    249.999954223633 1.71748220920563
    500.000091552734 1.92526078224182
    1000 1.84648036956787
    1999.99963378906 1.7127525806427
    4000.00073242188 1.51822936534882
    8000 1.20235741138458
    16000.0068359375 0.838871002197266
    };
    \addplot [thick, myRed, mark=oplus, forget plot, mark size=2.5pt]
    table {%
    125 1.60995066165924
    249.999954223633 1.76089072227478
    500.000091552734 1.8540506362915
    1000 1.84228694438934
    1999.99963378906 1.66812789440155
    4000.00073242188 1.50526607036591
    8000 1.18873643875122
    16000.0068359375 0.773829460144043
    };
    \addplot [thick, black, mark=diamond*, forget plot, mark size=2.5pt]
    table {%
    125 1.51870727539062
    249.999954223633 1.75816202163696
    500.000091552734 1.97833430767059
    1000 1.87417674064636
    1999.99963378906 1.77667593955994
    4000.00073242188 1.54895877838135
    8000 1.20225775241852
    16000.0068359375 0.819036483764648
    };
    
    \addplot [line width=1.25pt, black, mark=diamond*, mark size=4pt]
    table {%
    1 -1
    2 -1
    };
    \addlegendentry{Measurement}
    \addplot [line width=1.25pt, myLightBlue, mark=*, mark size=4pt]
    table {%
    1 -1
    2 -1
    };
    \addlegendentry{ARN (baseline)}
    \addplot [line width=1.25pt, myDarkBlue, mark=x, mark size=4pt]
    table {%
    1 -1
    2 -1
    };
    \addlegendentry{ARN (${K=1}$)}
    \addplot [line width=1.25pt, myGreen, mark=otimes, mark size=4pt]
    table {%
    1 -1
    2 -1
    };
    \addlegendentry{ARN (${K=2}$)}
    \addplot [line width=1.25pt, myOrange, mark=+, mark size=4pt]
    table {%
    1 -1
    2 -1
    };
    \addlegendentry{ARN (${K=1}$, spr.)}
    \addplot [line width=1.25pt, myRed, mark=oplus, mark size=4pt]
    table {%
    1 -1
    2 -1
    };
    \addlegendentry{ARN (${K=2}$, spr.)}
    
    \end{axis}
    
\end{tikzpicture}

%% file: figures/EDT_BRAS.tex
\begin{tikzpicture}
    \definecolor{myOrange}{RGB}{230, 159, 0}; 
    \definecolor{myLightBlue}{RGB}{86, 180, 233}; 
    \definecolor{myGreen}{RGB}{0, 158, 115}; 
    \definecolor{myYellow}{RGB}{240, 228, 66}; 
    \definecolor{myDarkBlue}{RGB}{0, 114, 178}; 
    \definecolor{myRed}{RGB}{213, 94, 0}; 
    \definecolor{myPurple}{RGB}{204, 121, 167}; 
    
    \begin{axis}[
    width=10cm,
    height=6cm,
    scale only axis,
    legend cell align={left},
    legend columns=2,
    legend style={
        at={(0.02, 0.02)},
        anchor=south west,
        fill opacity=0.7,
        draw opacity=1,
        text opacity=1,
        draw=lightgray
    },
    tick align=outside,
    tick pos=left,
    x grid style={darkgray},
    xmode=log,
    xlabel={Frequency [Hz]},
    xmin=125, xmax=16000,
    xtick style={color=black},
    xtick={125,250,500,1000,2000,4000,8000,16000},
    xticklabels={125,250,500,1k,2k,4k,8k,16k},
    y grid style={darkgray},
    ylabel={EDT [ms]},
    ymin=45, ymax=350,
    ytick style={color=black}
    ]
    
    \addplot [thick, myLightBlue, mark=*, forget plot, mark size=2.5pt]
    table {%
    125 167.851806640625
    249.999954223633 197.446960449219
    500.000091552734 252.231719970703
    1000 283.949249267578
    1999.99963378906 268.079742431641
    4000.00073242188 222.010681152344
    8000 129.210983276367
    16000.0068359375 50.6143913269043
    };
    \addplot [thick, myDarkBlue, mark=x, forget plot, mark size=2.5pt]
    table {%
    125 185.125564575195
    249.999954223633 232.338760375977
    500.000091552734 323.095397949219
    1000 291.704162597656
    1999.99963378906 282.214935302734
    4000.00073242188 236.337646484375
    8000 142.682525634766
    16000.0068359375 51.0836029052734
    };
    \addplot [thick, myGreen, mark=otimes, forget plot, mark size=2.5pt]
    table {%
    125 157.303344726562
    249.999954223633 230.110946655273
    500.000091552734 291.516418457031
    1000 286.241058349609
    1999.99963378906 255.913955688477
    4000.00073242188 222.25927734375
    8000 114.880683898926
    16000.0068359375 35.9918479919434
    };
    \addplot [thick, myOrange, mark=+, forget plot, mark size=2.5pt]
    table {%
    125 281.613739013672
    249.999954223633 282.420745849609
    500.000091552734 320.090362548828
    1000 300.297149658203
    1999.99963378906 260.128692626953
    4000.00073242188 231.979751586914
    8000 145.985931396484
    16000.0068359375 57.881908416748
    };
    \addplot [thick, myRed, mark=oplus, forget plot, mark size=2.5pt]
    table {%
    125 228.678955078125
    249.999954223633 220.442855834961
    500.000091552734 282.732482910156
    1000 263.916137695312
    1999.99963378906 275.161254882812
    4000.00073242188 239.545944213867
    8000 136.133834838867
    16000.0068359375 53.216968536377
    };
    \addplot [thick, black, mark=diamond*, forget plot, mark size=2.5pt]
    table {%
    125 235.895385742188
    249.999954223633 251.404754638672
    500.000091552734 308.066436767578
    1000 318.508270263672
    1999.99963378906 287.277923583984
    4000.00073242188 259.077697753906
    8000 164.404754638672
    16000.0068359375 83.397575378418
    };
    
    \addplot [line width=1.25pt, black, mark=diamond*, mark size=4pt]
    table {%
    1 -1
    2 -1
    };
    \addlegendentry{Measurement}
    \addplot [line width=1.25pt, myLightBlue, mark=*, mark size=4pt]
    table {%
    1 -1
    2 -1
    };
    \addlegendentry{ARN (baseline)}
    \addplot [line width=1.25pt, myDarkBlue, mark=x, mark size=4pt]
    table {%
    1 -1
    2 -1
    };
    \addlegendentry{ARN (${K=1}$)}
    \addplot [line width=1.25pt, myGreen, mark=otimes, mark size=4pt]
    table {%
    1 -1
    2 -1
    };
    \addlegendentry{ARN (${K=2}$)}
    \addplot [line width=1.25pt, myOrange, mark=+, mark size=4pt]
    table {%
    1 -1
    2 -1
    };
    \addlegendentry{ARN (${K=1}$, spr.)}
    \addplot [line width=1.25pt, myRed, mark=oplus, mark size=4pt]
    table {%
    1 -1
    2 -1
    };
    \addlegendentry{ARN (${K=2}$, spr.)}
    
    \end{axis}
    
\end{tikzpicture}

%% file: sections/conclusions.tex
\section{Conclusions}
\label{sec:conclusions}

This paper was concerned with approximating \acf{ART}, a room acoustic model based on \acf{GA}, using computationally efficient recursive networks of delay lines called room acoustic rendering networks.
Two network designs proposed by Bai \textit{et al}.~\cite{arn, fdn_art} served as starting point.
This paper proposed two key extensions.
The first consisted of a physically-informed design of the feedback matrices enabling explicit control of the scattering properties of wall materials.
The second involved high-order injection operators enabling to freely scale the number of early reflections modeled accurately and to improve the echo density buildup.

The resulting model was evaluated using objective measures of perceptual features, including frequency-dependent reverberation times, normalized echo density build-up, and early decay times.
The evaluation showed that together, the proposed extensions result in a significant improvement over the baseline.
This is especially so for rooms with non-convex geometries or unevenly distributed wall absorption, scenarios that are commonly encountered in practical settings and where scattering plays an important role.

The main venues for further improvement of the model are related to the requirements of unilossless matrix design, and to the computational complexity in cases where a high polygon count is required.
Future work will involve investigating how to further exploit the structure of \ac{ART} matrices to reduce approximation errors in the unilossless matrix optimization.
Further work will also explore complexity reduction strategies for (a) the operation of the recursive filter component, specifically in the case of real-time rendering of non-convex rooms, which require a fine spatial discretization, and (b) to reduce or remove the need for computationally costly ray-tracing in the precomputation stage, which can be problematic for applications where the room geometry is not known in advance, e.g.\@ \ac{AR}.
Finally, formal listening experiments will be carried out to validate the proposed model perceptually.

%% file: sections/appendix.tex
{\appendix[Geometry terms]

The definitions presented in this appendix can be found in the literature~\cite{rare}, but since we use different notation and nomenclature, we report them here for clarity.
The differential projected area $dA_{\text{p}\, x'}(x)$ and differential solid angle $d\omega_{x}(x')$, which appear in Eq.~(\ref{eq:radiance definition integral}), are defined as:
\begin{align}
    dA_{\text{p}\, x'}(x)
    = & \ 
    \max(0, n(x) \cdot v(x|x'))
    dA(x)
    \, ,
    \label{eq:diff proj area}
    \\
    d\omega_{x}(x')
    = & \ 
    \frac{
    \max(0, n(x') \cdot v(x'|x))
    }{\Lnorm{x - x'}^2}
    dA(x')
    \, ,
    \label{eq:diff solid angle}
\end{align}
where $n(x)$ is the surface normal at point $x$, and $\Lnorm{x - x'}$ is the distance between $x$ and $x'$.
Note that ${n(x') \cdot v(x'|x)}$ is the cosine of the angle of incidence from $x$ onto $dA(x')$, and that applying $\max(0, \cos)$ means setting the cosine value to 0 if the direction $v(x'|x)$ is on the ``back side'' of $dA(x')$.

The following definitions are for the terms which appear in Eq.~(\ref{eq:refl kernel}), making up the reflection kernel $R \Lpar{x, x', v(x'|x'')}$.
The term $\rho$ is the so-called \acf{BRDF}, defined as
\begin{align}
    \rho \Lpar{x'|\, v(x'|x), v(x'|x'')}
    = & \ \nonumber
    \frac{
        1
    }{
        (n(x') \cdot v(x'|x))
        L_\text{in} \Lpar{x', v(x'|x)}
    }
    \\ & \ 
    \times
    \frac{
        \partial L_\text{out} \Lpar{x', v(x'|x'')}
    }{
        \partial \omega_{x'}(x)
    }
    \, .
    \label{eq:BRDF definition}
\end{align}
The term $G$ is defined as
\begin{align}
    G(x, x')
    = & \ 
    \frac{
        \max(0, n(x) \cdot v(x|x'))
        \max(0, n(x') \cdot v(x'|x))
    }{\Lnorm{x - x'}^2}
    \, .
    \label{eq:geom term}
\end{align}
Note how ${G(x, x') = G(x', x)}$, and
\begin{align}
    G(x, x')
    dA(x)
    dA(x')
    = & \ 
    dA_{\text{p}\, x'}(x)
    d\omega_{x}(x')
    \\
    = & \ 
    dA_{\text{p}\, x}(x')
    d\omega_{x'}(x)
    \, .
    \label{eq:etendue}
\end{align}
The term ${V(x, x')}$ is a visibility operator, equal to 1 if there is an unobstructed line of sight between the points $x$ and $x'$ and 0 otherwise.
The term ${D(x, x')}$ is a delay operator, which may also include energy absorption due to propagation.

In the case of a point source located in $x_s$, the primary reflected radiance is
\begin{align}
    L_0 \Lpar{x, v(x|x'), t}
    = & \ \nonumber
    \rho \Lpar{x|\, v(x|x_s), v(x|x')}
    V(x_s, x)
    \\ & \ 
    \times
    D(x_s, x)
    \Delta_s(x)
    \frac{
        P_s(t)
    }{
        4\pi
    }
    \frac{
        d\omega_{x_s}(x)
    }{
        dA(x)
    }
    \, ,
    \label{eq:primary reflected radiance}
\end{align}
where ${P_s(t)}$ is the source power, and ${\Delta_s(x)}$ is the source's transfer function (e.g.\@ loudspeaker's polar pattern) towards $x$.

The sound intensity detected by a point receiver located in $x_r$ is defined by the sum ${I(x_r, t) = I_D(x_r, t) + I_L(x_r, t)}$ of the direct-incidence component $I_D(x_r, t)$ and the reflected component $I_L(x_r, t)$, defined as
\begin{align}
    I_D(x_r, t) & =
    \Delta_r(x_s)
    V(x_s, x_r)
    D(x_s, x_r)
    \frac{
        \Delta_s(x_r)
        P_s(t)
    }{
        4\pi \Lnorm{x_r - x_s}^2
    }
    \label{eq:direct detection}
    \, ,
    \\ \nonumber
    I_L(x_r, t) & =
    \iint\displaylimits_A
    \Delta_r(x)
    V(x, x_r)
    D(x, x_r)
    \\
    & \hphantom{= \iint_\Omega.} \times
    L \Lpar{x, v(x|x_r), t}
    d\omega_{x_r}(x)
    \, ,
    \label{eq:scattered detection}
\end{align}
where ${\Delta_r(x)}$ is the receiver's transfer function (e.g.\@ a microphone polar pattern, or \ac{HRTF}) towards $x$.

}

%% file: biography.tex
\vspace{-33pt}
\begin{IEEEbiography}[{\includegraphics[width=1in,height=1.25in,clip,keepaspectratio]{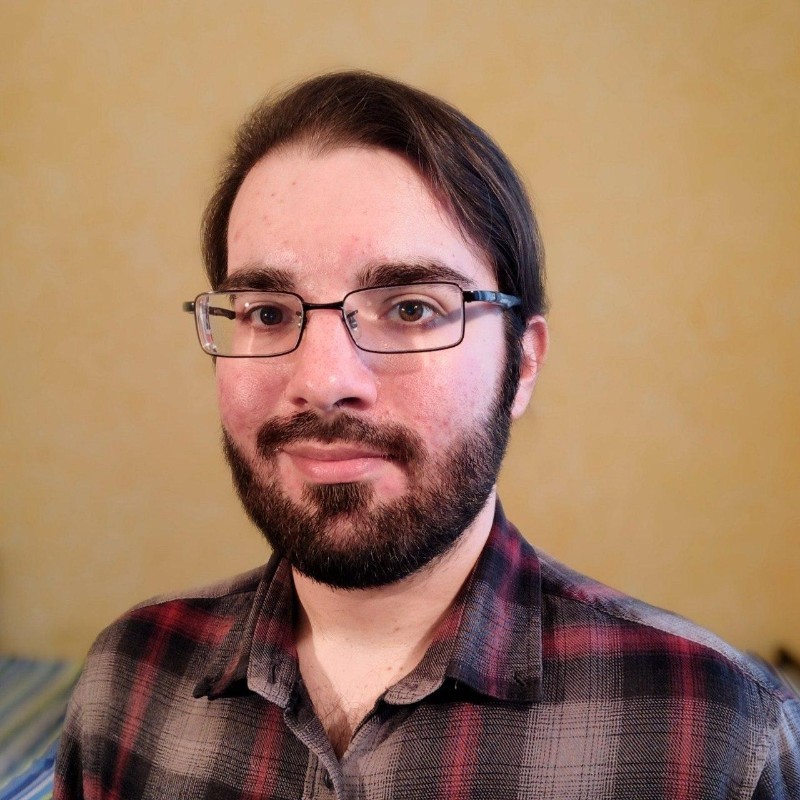}}]{Matteo Scerbo}
Matteo Scerbo received his Bachelor's degree in computer engineering from the Università di Genova (UniGe) in 2017.
He then specialized in acoustics and audio signal processing, receiving his Master's degree in Music and Acoustics Engineering at the Politecnico di Milano (PoliMi) in 2021.
Since 2021, he has been a PhD student in Audio Engineering at the University of Surrey (UoS), investigating efficient methods for room acoustics modeling.
His interests include room acoustics modeling, digital signal processing, and virtual reality.
\end{IEEEbiography}
\vspace{-33pt}
\begin{IEEEbiography}[{\includegraphics[width=1in,height=1.25in,clip,keepaspectratio]{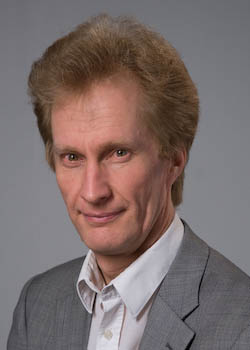}}]{Lauri Savioja}
Lauri Savioja (M'00-SM'08) works as a professor at the Department of Computer Science at Aalto University, Finland. He received his doctorate from the Helsinki University of Technology in 1999 majoring in computer science while his thesis topic was room acoustic modeling. His research interests include room acoustics, virtual reality, and parallel computation. Prof. Savioja is a fellow of the Audio Engineering Society (AES) and of the Acoustical Society of America (ASA), senior member of the Institute of Electrical and Electronics Engineers (IEEE), and a life member of the Acoustical Society of Finland.
\end{IEEEbiography}
\vspace{-33pt}
\begin{IEEEbiography}[{\includegraphics[width=1in,height=1.25in,clip,keepaspectratio]{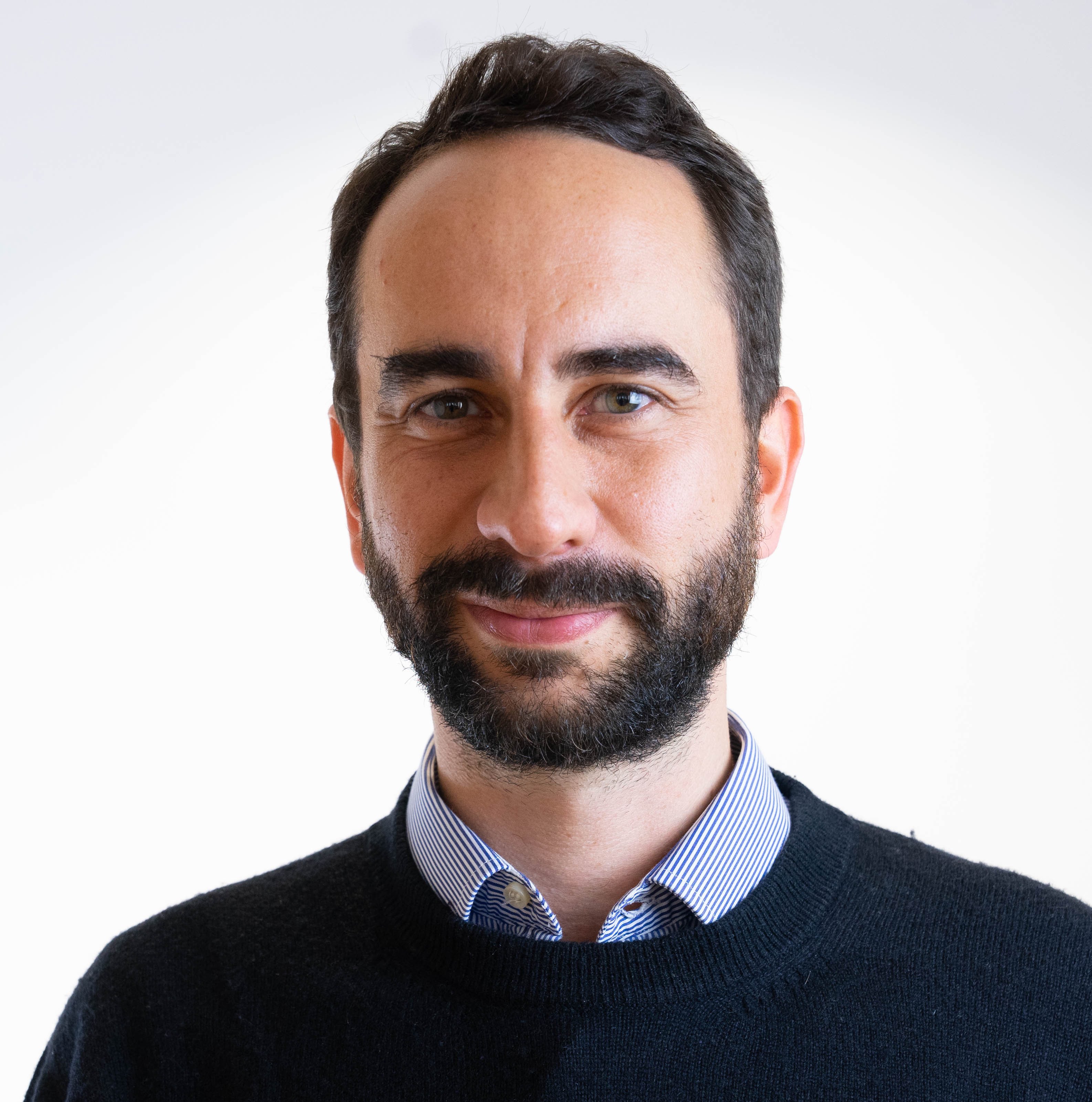}}]{Enzo De Sena}
Enzo De Sena (Senior Member, IEEE) received the
M.Sc. degree (cum laude) in telecommunication engineering from the Università di Napoli ``Federico II,'' Naples, Italy, in 2009, and the Ph.D. degree in electronic engineering from King’s College London, London, U.K., in 2013. 
He is currently an Associate Professor (Reader) at the University of Surrey, Guildford, U.K., where he serves as Director of the Institute of Sound Recording (IoSR). 
Between 2013 and 2016, he was a Postdoctoral Researcher with KU Leuven, Leuven, Belgium. He held visiting Researcher positions at Stanford University, Stanford, CA, USA, Aalborg University, Aalborg, Denmark, Imperial College London, London, U.K., and King’s College London, London, U.K. 
His research interests include room acoustics modelling, sound field reproduction, beamforming, and binaural modeling. 
He is a Member of the IEEE Audio and
Acoustic Signal Processing Technical Committee, and an Associate Editor for the EURASIP Journal on Audio, Speech, and Music Processing and IEEE/ACM TRANSACTIONS ON AUDIO SPEECH AND LANGUAGE PROCESSING. 
He was the recipient of an EPSRC New Investigator Award and co-recipient of Best Paper Awards at WASPAA-21 and AVAR-22. For more information see: desena.org.
\end{IEEEbiography}